\documentclass[11pt,a4paper]{article}
\usepackage{epsfig,graphicx,amsfonts,amsmath,amssymb}

\usepackage{cite}
\RequirePackage{mathrsfs}

\newlength{\dinwidth}
\newlength{\dinmargin}
\def\fig#1{{Fig.~(\ref{#1})}}
\def\eq#1{{Eq.~(\ref{#1})}}
\newcommand{\Le}{\left(}
\newcommand{\Ra}{\right)}

\newcommand{\beq}{\begin{equation}}
\newcommand{\eeq}{\end{equation}}
\newcommand{\beqar}{\begin{eqnarray}}
\newcommand{\eeqar}{\end{eqnarray}}
\newcommand{\D}{\partial}

\newcommand{\ep}{\varepsilon}

\newcommand{\tv}{\textsl{v}}

\newcommand{\T}{{\cal T}}
 
\date{}
\begin{document}

\title {{~}\\
{\Large \bf  On QCD RFT corrections to the propagator of reggeized gluons }}
\author{ 
{~}\\
{\large 
S.~Bondarenko$^{(1) }$,
S.~Pozdnyakov$^{(1) }$
}\\[7mm]
{\it\normalsize  $^{(1) }$ Physics Department, Ariel University, Ariel 40700, Israel}\\
}

\maketitle
\thispagestyle{empty}

\begin{abstract}
  
\end{abstract}

 We calculate an one loop QCD Regge Field Theory (RFT) correction to the propagator of reggeized gluons basing on the  QCD effective action  of Lipatov,
\cite{LipatovEff,LipatovEff1,Our1,Our2,Our3,Our4}, and results of \cite{Our5}
where Dyson-Schwinger hierarchy of the equations for the correlators of reggeized
gluon fields was derived. The correction is calculated entirely in the framework of RFT with the use of the obtained expressions for the RFT bare triple Reggeon vertices and 
propagator of reggeized gluons, the cases of bare propagator and propagator calculated to one-loop precision are considered separately.
In both results the obtained correction represents non-eikonal contributions to the propagator kinematically suppressed by $1/\sqrt{s}$ factor in comparison to the usual LLA contributions.
The further application of the obtained results is discussed as well.

\section{Introduction}

 $\,\,\,\,\,\,$ The action for an interaction of the reggeized gluons, introduced in \cite{LipatovEff,LipatovEff1},  see also \cite{Our1,Our2,Our3,Our4},
describes quasi-elastic amplitudes of the high-energy scattering processes
in the  multi-Regge kinematics. The applications of this action for the description of the high energy processes
and calculation of sub-leading, unitarity  corrections to the
amplitudes and production vertices can be found in \cite{EffAct}, see also \cite{Fadin}.
The generalization of the formalism for a purpose of the calculation of production amplitudes and impact factors was obtained in \cite{Our3},
where the prescription of the calculation of $S$-matrix elements of the different processes was given accordingly to an approach of \cite{Faddeev}.
This effective action formalism, based on the reggeized gluons as main degrees of freedom, see \cite{BFKL}, can be considered as a reformulation of the RFT (Regge Field Theory) calculus introduced
in \cite{Gribov}, see also \cite{0dim1,0dim2,0dim3,0dim4,Pom1,Pom2,Pom22,Pom3,Pom4}, for the case of high energy QCD.
It was underlined in \cite{LipatovEff,LipatovEff1} that the main purposes of the approach is the construction of the
$S$-matrix unitarity in the direct and crossing channels of the scattering processes through the multi-Reggeon dynamics described by the  vertices  of multi-Reggeon
interactions, see other and simirar approaches in \cite{Venug,Kovner,Hatta1,BKP,GLR,BK,TripleV,Hetch}. The unitarity of the Lipatov's formalism, therefore, is related to 
the unitary corrections in both RFT and QCD sectors of the theory.

 Similarly to the phenomenological theories of interacting Reggeons, see \cite{Gribov,0dim1,0dim2,0dim3,0dim4} and references therein, there is a very natural question to ask: what are the corrections 
to the amplitudes which come from the pure RFT sector of the formalism. Namely, consider the Lipatov's effective action
for reggeized gluons $A_{\pm}$, formulated as  RFT (Regge Field Theory) which can be obtained by an integration out the gluon fields $v$ in the
generating functional for the $S_{eff}[v,\,A]$:
\beq\label{Add1}
e^{\imath\,\Gamma[A]}\,=\,\int\,D v\,e^{\imath\,S_{eff}[v,\,A\,]}
\eeq
where
\beq\label{Add2}
S_{eff}\,=\,-\,\int\,d^{4}\,x\,\Le\,\frac{1}{4}\,G_{\mu \nu}^{a}\,G^{\mu \nu}_{a}\, \,+\,tr\,\left[\,\Le\,\T_{+}(v_{+})\,-\,A_{+}\,\Ra\,j_{reg}^{+}\,+\,
\Le\,\T_{-}(\textsl{v}_{-})\,-\,A_{-}\,\Ra\,j_{reg}^{-}\,\right]\,\Ra\,,
\eeq
with
\beq\label{Add3}
\T_{\pm}(v_{\pm})\,=\,\frac{1}{g}\,\D_{\pm}\,O(v_{\pm})\,=\,
v_{\pm}\,O(v_{\pm})\,,\,\,\,\,
j_{reg\,a}^{\pm}\,=\,\frac{1}{C(R)}\,\D_{i}^{2}\,A_{a}^{\pm}\,,
\eeq
here  $C(R)$ is eigenvalue of Casimir operator in the representation R, $tr(T^{a} T^{b})\,=\,C(R)\,\delta^{a b}$\, see \cite{LipatovEff,LipatovEff1,Our1,Our2}.
The form of the Lipatov's operator  $O$ (and correspondingly $\T$) depends on the particular process of interests, see \cite{Our4},
we take it in the form of  the Wilson line (ordered
exponential) for the longitudinal gluon fields in the adjoint representation:
\beq\label{Add4}
O(v_{\pm})\,=\,P\,e^{g\,\int_{-\infty}^{x^{\pm}}\,d x^{\pm}\,v_{\pm}(x^{+},\,x^{-},\,x_{\bot})}\,,\,\,\,\,\tv_{\pm}\,=\,\imath\,T^{a}\,v_{\pm}^{a}\,,
\eeq
see also \cite{Nefedov}. 
There are additional kinematical constraints for the reggeon fields
\beq\label{Ef4}
\partial_{-}\,A_{+}\,=\,\partial_{+}\,A_{-}\,=\,0\,,
\eeq
corresponding to the strong-ordering of the Sudakov components in the multi-Regge kinematics,
see \cite{LipatovEff,LipatovEff1,Our4}. The action is constructed by the request that the  LO value of the classical gluon fields in the solutions of equations of motion will be fixed as
\beq\label{Ef8}
\textsl{v}_{\pm}\,=\,A_{\pm}\,.
\eeq
In the light-cone gauge $\textsl{v}_{-}\,=\,0$,  the equations of motion can be solved and the general expressions for the gluon fields can be written in the following form:
\beq\label{Ef9}
v_{i}^{a}\,\rightarrow\,v^{a}_{i\,cl}(A_{\pm})\,+\,\ep_{i}^{a}\,,\,\,\,\,v_{+}^{a}\,\rightarrow\,v_{+\,cl}^{a}(A_{\pm})\,+\,\ep_{+}^{a}\,.
\eeq
The integration in respect to the fluctuations around the classical solutions provides QCD loop corrections to the effective vertices of the Lipatov's action which now can be written as functional of 
the Reggeon fields only\footnote{In order to make the notations shorter, we change the position of the color and other indexes of the vertices further in the article, preserving only the
overall number of the indexes.} :
\beq\label{Sec1}
\Gamma = \sum_{n,m\,=\,1}\Le\,A_{+}^{\,a_1}\cdots A_{+}^{\,a_n} \Le K^{+\,\cdots\,+}_{-\,\cdots\,-}\Ra^{a_1\cdots\, a_n}_{b_1\cdots\, b_m} A_{-}^{\,b_1} \cdots\,A_{-}^{\,b_m} \Ra = 
-\,A_{+\,x}^{\,a} \D_{i}^{2}\,A_{-\,x}^{\,a} + A_{+\,x}^{\,a} \Le K_{x y}^{\,a\, b} \Ra^{+}_{-} A_{-\,y}^{\,b} + \cdots\,,
\eeq
in general  the summation 
on the color indexes in the r.h.s of the equation means the integration on the corresponding coordinates as well.
Now we see, that the theory have two different sources of any perturbative/unitary corrections. The first one comes from the QCD side of the framework
 and based on the precision of the effective vertices (kernels)
calculated in the pure QCD. Another source of the corrections is described by the diagrams directly from the RFT sector of the theory constructed entirely in terms of the Reggeon fields and 
\eq{Sec1} vertices to some given QCD precision. These RFT type of corrections were calculated a 
lot in the previous phenomenological RFT theories, see \cite{0dim1,0dim2,0dim3,0dim4} and references therein for example; there are the QCD RFT corrections to the propagator
and vertices of the \eq{Sec1} action as well. In the paper \cite{Our5} the Dyson-Schwinger hierarchy of the equations for the correlators of reggeized gluons was derived in the formalism, of 
that allows to define and calculate these RFT sector corrections to any correlator of interests. We also note, that formally this hierarchy is similar to the Balitsky hierarchy
of equations and BK-JIMWLK approaches , see \cite{Venug,Kovner,Hatta1,BK}, and there is a correspondence between different degrees of freedom such as  reggeized gluons and Wilson line operators, 
see details in \cite{Our5,Hetch}.

 In this paper we calculate a one Reggeon loop correction to the propagator of reggeized gluons. In order to perform the calculations we 
need the expressions for the vertices of interactions of three Reggeons. These vertices are calculated as well to the bare QCD precision and basing on this result
we construct the RFT Reggeon loop contribution to the propagator using \cite{Our5} equation for the correlator of two Reggeon fields. 
We separately consider two different forms of the Reggeon loop. The first one we construct with the use of the only bare QCD vertices in the equations of hierarchy, whereas
in the second case we account also a leading order one-loop QCD correction to the correlators that leads to the different final expressions in two cases. 
In both answers, nevertheless, the obtained expressions provide non-zero amplitudes only in convolution with non-eikonal impact factors introduced and the final answers are 
suppressed in comparison to the LLA contributions of the similar order by the additional $1/\sqrt{s}$ factor. 
Therefore, the paper is organized as follows. In the next section we remind some basic definitions from the \cite{Our2}. The Section 3 is dedicated to the calculation of the one loop QCD
propagator of the reggeized gluons. The Section 4 is about the calculations of the bare QCD  vertices of the triple Reggeon interactions required for the one RFT loop construction.
Sections 5 and 6 are about the derivation of the expression for the two Reggeon fields correlator to one RFT loop precision and it's calculation. The last Section is the Conclusion of the article, there are also Appendixes where some technical details of the calculations are present.

\section{One loop effective action}

This section is based mainly on the results from \cite{Our2} paper, therefore we remind shortly only the some important formulas from there.
In the derivation of the QCD RFT action the following representation of the gluon fields is used:
\beq\label{Ef1}
v_{i}^{a}\,\rightarrow\,v^{a}_{i\,cl}\,+\,\ep_{i}^{a}\,,\,\,\,\,v_{+}^{a}\,\rightarrow\,v_{+\,cl}^{a}\,+\,\ep_{+}^{a}\,,
\eeq
at the next step we expand the Lagrangian of the effective action around the classical solutions. Preserving in the expression only terms which are quadratic with respect to the fluctuation fields,  
we obtain for this part of the action:
\beqar\label{Ef2}
S_{\ep^{2}}& = &-\frac{1}{2}\,\int\,d^{4} x\,\left.\Big( \ep_{i}^{a}\Le \delta_{ac}\Le \delta_{ij}\,\Box\, +\D_{i}\,\D_{j} \Ra - \right.\right. \nonumber \\
&-&
2 g f_{abc} \Le \delta_{ij} \,v_{k}^{b\,cl} \D_{k}  -
2\, v_{j}^{b\,cl} \D_{i}  + v_{i}^{b\,cl} \D_{j} - \delta_{ij} \,v_{+}^{b\,cl} \D_{-}  \Ra    -\nonumber\\
&-&\left.\,
\,g^{2}\,f_{a b c_{1}}\,f_{c_{1} b_{1} c}\,
\Le \delta_{i j}\,v_{k}^{b\,cl}\,v_{k}^{b_{1}\,cl}\,
-\,v_{i}^{b_1\,cl}\,v_{j}^{b\,cl}\,\Ra\,\Ra\,\ep_{j}^{c}\,+\,\nonumber\\
&+&\,
\,\ep_{+}^{a}\Le -2 \,\delta^{a c}\,\D_{-}\D_{i} -2 g f_{abc} \Le  v_{i}^{b\,cl} \D_{-} - \Le \D_{-} v_{i}^{b\,cl}\Ra \Ra \Ra\,\ep_{i}^{c}\,
+\,\nonumber\\
&+&\,\left.\,
\ep_{+}^{a}\, \delta_{a c}\, \D_{-}^{2} \,\ep_{+}^{c} -
\,g\,\ep_{+\,x}^{a}\,\int\,d^{4} y\,\Le U^{a\,b\,c}_{1} \Ra^{+}_{x\,y} \Le \D_{i} \D_{-} \rho_{b}^{i}\Ra_{x}\,\ep_{+\,y}^{c}\,\Ra
\,=\,\nonumber \\
&=&\,-\,\frac{1}{2}\,\ep_{\mu}^{a}\,\Le\,\Le M_{0} \Ra_{\mu\, \nu}^{ac}\,+\,\Le M_{1} \Ra_{\mu\, \nu}^{ac}\,+\,
\Le M_{2} \Ra_{\mu\, \nu}^{ac}\,+\,\Le M_{L} \Ra_{\mu\, \nu}^{ac}\Ra\,\ep_{\nu}^{c}\,.
\eeqar
Here we defined $\Le M_{i} \Ra_{\mu\, \nu}^{ac}\,\propto\,g^{i}$ and note that
\beq\label{Ef3}
\Le M_{1}\Ra_{-\,i}\,=\,-\, g f_{abc} \Le  v_{i}^{b\,cl} \,\overrightarrow{\D_{-}} - \Le \D_{-} v_{i}^{b\,cl}\Ra \Ra\,,\,\,\,
\Le M_{1}\Ra_{i\,-}\,=\,-\, g f_{abc} \Le  \,\overleftarrow{\D_{-}}\,  v_{i}^{b\,cl}- \Le \D_{-} v_{i}^{b\,cl}\Ra \Ra\,.
\eeq
The last term in \eq{Ef2} expression, denoted as $\Le M_{L} \Ra_{\mu\, \nu}^{ac}$ represents contribution of the Lipatov's effective current into the action. 
This term is defined trough the following function:
\beq\label{Ef444}
\Le U_{1}^{a\, b\, c} \Ra_{x y}^{+}\,=\,
tr[\,f_{a}\,G^{+}_{x y}\,f_{c}\,O_{y}\,f_{b}\,O^{T}_{x}]\,+\,
tr[\,f_{c}\,G^{+}_{y x}\,f_{a}\,O_{x}\,f_{b}\,O^{T}_{y}]\,,
\eeq
see \cite{Our2} and Appendix B. 
There is also
some color density function introduced as 
\beq\label{Ef31}
\D_{i}\,\D_{-}\,\rho_{a}^{i}\,=\,-\,\frac{1}{N}\,\D_{\bot}^{2}\,A_{a}^{+}\,,
\eeq
or
\beq\label{Ef32}
\rho_{a}^{i}\,=\,\frac{1}{N}\,\D_{-}^{-1}\,\Le\D^{i}\,A_{-}^{a}\Ra\,,
\eeq
see \cite{Our4,Venug,Kovner,Hatta1}.
Now we can perform the integration obtaining the one loop QCD effective action:
\beqar\label{Ef5}
\Gamma\,& = &\,\int\,d^{4} x\,\Le L_{YM}(v_{i}^{cl},\, v_{+}^{cl})- v_{+\,cl}^{a}\,J_{a}^{+}(v_{+}^{cl})- A_{+}^{a}\,\Le\,\D_{i}^{2}\,A_{-}^{a}\,\Ra\,\Ra + \nonumber \\
& + &\,
\frac{\imath}{2}\,Tr\,\ln\Le\,\delta_{\rho\,\nu}\,+\,G_{0\,\rho\,\mu}\,\Le\, \Le M_{1}\Ra_{\mu\,\nu}\,+\,\Le M_{2}\Ra_{\mu\,\nu}\,+\,
\Le M_{L}\Ra_{\mu\,\nu}\,\Ra\,\Ra\, + \nonumber \\
&+&\,\frac{1}{2}\,\int\,d^{4} x\,\int\,d^{4} y\,
j_{\mu\,x}^{\,a}\,G_{\mu\,\nu}^{\,a b}(x,y)\,j_{\nu\,y}^{\,b}\,.
\eeqar
Here we have $\,G_{\,0\,\nu\,\mu}$ as bare  gluon propagator
\beq\label{Ef6}
\Le M_0\Ra_{\,\mu\,\nu}\,G_{\,0\,\nu\,\rho}\,=\,\delta_{\mu \rho}\,,
\eeq
see Appendix A; the full gluon propagator is defined as
\beq\label{Ef7}
G_{\mu \nu}^{ac}\,=\,\left[\,\Le M_{0} \Ra_{\mu \nu}^{ac}\,+\,\Le M_{1} \Ra_{\mu \nu}^{ac}\,+\,\Le M_{2} \Ra_{\mu \nu}^{ac}\,+\,\Le M_{L}\Ra_{\mu\,\nu}^{ac}\,\right]^{-1}\,
\eeq
and can be written in the form of the following  perturbative series:
\beq\label{Eff8}
G_{\mu \nu}^{ac}(x,y)\,=\,G_{0\,\mu \nu}^{ac}(x,y)\,-\,\int\,d^4 z\,G_{0\,\mu \rho}^{ab}(x,z)\,
\Le \,\Le M_{1}(z)\Ra_{\rho \gamma}^{bd}\,+\,\Le M_{2}(z)\Ra_{\rho \gamma}^{bd}\,+\,\Le M_{L}(z)\Ra_{\rho\,\gamma}^{bd}\, \Ra G_{\gamma \mu}^{dc}(z,y)\,;
\eeq
the auxiliary currents $j_{\mu\,x}^{\,a}$ and  $j_{\nu\,y}^{\,b}$ are requested for the many-loops calculations of the effective action, in our case 
of calculation to one loop QCD precision we take them  zero from the beginning.

\section{Propagator of reggeized gluons}

 In order to calibrate the calculations and introduce some useful further notations, in this Section we rederive the calculations of the propagator of reggeized gluons to one loop QCD precision
done in \cite{Our2}.  
The interaction of reggeized gluons $A_{+}$ and $A_{-}$ is defined trough an effective vertex of the interaction in \eq{Ef5} as:
\beq\label{Pro1}
\Le\,K^{a\,b}_{x\,y}\,\Ra^{+\,-}\,=\,K^{a\,b}_{x\,y}\,=\,\Le\,\frac{\delta^{2}\,\Gamma}{\delta A_{+\,x}^{a}\,\delta A_{-\,y}^{b}}\,\Ra_{A_{+},\,A_{-},\,v_{f\,\bot}\,=\,0}\,,
\eeq
Due the properties of the Reggeon fields, see \eq{Ef4}, this vertex in the expressions we obtain in the following form:
\beq\label{Pro11}
K^{a\,b}_{x\,y}\,=\,K^{a\,b}(x^{+},\,x_{\bot}\,;y^{-},\,y_{\bot})\,=\,\int d x^{-}\,d y^{+}\,\tilde{K}^{a b}\Le x^{+},\,x^{-},\,x_{\bot}\,;\,y^{+},\,y^{-},\,y_{\bot}  \Ra\,.
\eeq
At high-energy approximation, the transverse coordinates are factorized from the longitudinal ones in the LO and NLO vertices of $A_-$ and $A_+$ Reggeon fields interactions:
\beq\label{Pro12}
\tilde{K}^{a b}\Le x^{+},\,x^{-},\,x_{\bot}\,;\,y^{+},\,y^{-},\,y_{\bot}  \Ra\,=\,\delta(y^{-}\,-\,x^{-})\,\delta(x^{+}\,-\,y^{+})\,K^{a\,b}(x_{\bot}\,,y_{\bot})\,.
\eeq
Indeed, the LO vertex in the formalism has the following form:
\beq\label{Pro5}
 K^{a\,b}_{x y\,0}\,=\,-\,\delta^{a\,b}\,\delta_{x_{\bot}\,y_{\bot}}\,\D_{i\,x}^{2}\,
\eeq
and the similar property is holding for the NLO vertex, see Appendix C.
The bare propagator of the action for the Reggeon can be  defined similarly:
\beq\label{Pro21}
\,D^{a b}_{z y\,0}\,=\,D^{a b}_{+ -\,0}(x^{+},\,x_{\bot}\,;y^{-},\,y_{\bot})\,=\,
\int d x^{-}\,d y^{+}\,\tilde{D}^{a b}_{+ -\,0}\Le x^{+},\,x^{-},\,x_{\bot}\,;\,y^{+},\,y^{-},\,y_{\bot}  \Ra\,
\eeq
with
\beq\label{Pro22}
\tilde{D}^{a b}_{+ -\,0}\Le x^{+},\,x^{-},\,x_{\bot}\,;\,y^{+},\,y^{-},\,y_{\bot}  \Ra\,=\,\delta(y^{-}\,-\,x^{-})\,\delta(x^{+}\,-\,y^{+})\,
\,D_{+ -\,0}^{a b}(x_{\bot},\,y_{\bot})\,.
\eeq
The propagator satisfies the following equation:
\beq\label{Pro2}
\int\,d^{4} z\,\Le\,\tilde{K}^{a b}_{x z\,0}\Ra^{-\,+}\,\Le\,\tilde{D}^{b c}_{z y\,0}\Ra_{+\,-}\,=\,\delta^{a c}\,\delta^{4}_{x y}\,
\eeq
that provides
\beq\label{Pro61}
D_{+ -\, 0}^{a b}(x_{\bot},\,y_{\bot})\,=\,D_{0}^{a b}(x_{\bot},\,y_{\bot})\,=\,\delta^{a b}\,\int\,\frac{d^{2} p}{(2\pi)^2}\frac{e^{-\imath\,p_{i}\,(x^{i}\,-\,y^{i})}}{p_{\bot}^{2}}\,.
\eeq
Considering, correspondingly, the perturbative expansion of the kernel
\beq\label{Pro3}
 K^{b d}_{z w}\,=\,\sum_{k\,=\,0}\, K^{b d}_{z w\,k}\,,
\eeq
the full propagator of reggeized gluons is defined in the form of the perturbative series as well:
\beq\label{Pro6}
 D_{x y}^{a c}\,=\, D_{x y\,0}^{a c}\,-\,\int\,d^{4} z\,\int\,d^{4} w\, D_{x z\,0}^{a b}\,
\Le\,\sum_{k\,=\,1}\, K^{b d}_{z w\,k}\,\Ra\, D_{w y}^{d c}\,.
\eeq
To the leading order precision we need to calculate the $K^{b d}_{z w\,1}$ kernel, the calculations are presented in Appendix C. Therefore, we obtain to this order:
\beq\label{Pro62}
 D_{x y}^{a c}\,=\, D_{x y\,0}^{a c}\,-\,\int\,d^{4} z\,\int\,d^{4} w\,\Le \D_{i\,z}^{2} D_{x z\,0}^{a b}\,\Ra\,
K^{b d}_{z w\,1}\, D_{w y}^{d c}\,.
\eeq
Introducing
\beq\label{Pro7}
 D_{x y}^{a c}\,=\,\delta^{a c}\delta(y^{-}\,-\,x^{-})\,\delta(x^{+}\,-\,y^{+})\,\int\,\frac{d^2 p}{(2 \pi)^2}\,\tilde{D}(p_{\bot},\,\eta)\,
e^{-\imath\,p_{i}\,\Le x^{i} - y^{i}\Ra}\,,
\eeq
we obtain finally:
\beq\label{Pro8}
\tilde{D}^{a b}(p_{\bot},\eta)\,=\,\frac{\delta^{a b}}{p_{\bot}^{2}}\,+\,
\epsilon(p_{\bot}^{2})\,\int^{\eta}_{0}\,d\eta^{'}\,\tilde{D}^{a b}(p_{\bot},\eta^{'})\,
\eeq
with
\beq\label{Pro9}
\epsilon(p_{\bot}^{2})\,=\,-\,\frac{\alpha_{s}\,N}{4\,\pi^{2}}\,\int\,d^{2} k_{\bot}\,
\frac{ \,p_{\bot}^{2}}{k_{\bot}^{2}\,\Le\,p_{\bot}\,-\,k_{\bot}\,\Ra^{2}}\,
\eeq
as trajectory of the propagator of reggeized gluons. Rewriting this equation as the differential one:
\beq\label{Pro10}
\frac{\D\,\tilde{D}^{a b}(p_{\bot},\eta)}{\D\,\eta}\,=\,\tilde{D}^{a b}(p_{\bot},\eta)\,\epsilon(p_{\bot}^{2})\,
\eeq
we obtain the final expression for the propagator:
\beq\label{Pro1111}
\tilde{D}^{a b}(p_{\bot},\eta)\,=\,\frac{\delta^{a b}}{p_{\bot}^{2}}\,e^{\,\eta\,\epsilon(p_{\bot}^{2})}\,,
\eeq
with $\eta$ defined in some rapidity interval $0\,<\,\eta\,<\,Y\,=\,\ln(s/s_{0})$ of interest; of course it is the BFKL propagator for reggeized gluons calculated to one loop QCD precision, 
see \cite{BFKL}.

\section{Bare vertices of triple Reggeons interactions}

The first contribution to any bare (QCD zero-loop) vertex of the three Reggeon fields interactions  is coming from the  Yang-Mills action and it is defined as following:
\beq\label{BV1}
\Le K^{a b c }_{x y z}\Ra_{0\,YM}^{\mu \nu \rho}\,=\,\int\, d^{4} w\, \Le\,\frac{\delta^{3}\,L_{YM}(v_{i}^{cl}(A),\,v_{+}^{cl}(A))}{\delta A_{\mu}^{a}(x)\,\delta A_{\nu}^{b}(y)\, \delta A_{\rho}^{c}(z) }\,\Ra_{A\,=\,0}\,,\,\,\,\,\,\,\mu \nu \rho=(+,\,-)\,
\eeq
with the QCD Lagrangian in light-cone gauge
\beq\label{BV2}
L\,=\,-\,\frac{1}{4}\,F_{ij}^{a}\,F_{ij}^{a}\,+\,F_{i+}^{a}\,F_{i-}^{a}\,+\,\frac{1}{2}\,F_{+-}^{a}\,F_{+-}^{a}\,,
\eeq
where $\,F_{+-}^{a}\,F_{+-}^{a}\,$ term does not consists  transverse fields and $\,F_{ij}^{a}\,F_{ij}^{a}\,$ term does not consist longitudinal fields. 
Correspondingly, we have for the non-zero contributions in \eq{BV1}:
\beqar\label{BV3}
&\,&\frac{\delta^{3}\,L_{YM}(v_{i}^{cl},\,v_{+}^{cl})}{\delta A_{\mu}^{a}(x)\,\delta A_{\nu}^{b}(y)\, \delta A_{\rho}^{c}(z) }\,= \,
\frac{\delta^{3}\,L_{YM}}{\delta v_{\mu_1}^{cl\,a_1}\,\delta v_{\mu_2}^{cl\,a_2}\, \delta v_{\mu_3}^{cl\,a_3}}\,\frac{\delta\,v_{\mu_1}^{cl\,a_1} }{\delta A_{\mu}^{a}}\,
\frac{\delta\,v_{\mu_2}^{cl\,a_2} }{\delta A_{\nu}^{b}}\,\frac{\delta\,v_{\mu_3}^{cl\,a_3} }{\delta A_{\rho}^{c}}\,+\,\nonumber \\
&+&\,
\frac{\delta^{2}\,L_{YM}}{\delta v_{\mu_1}^{cl\,a_1}\,\delta v_{\mu_2}^{cl\,a_2}}\,\Le\,
\frac{\delta^2 v_{\mu_1}^{cl\,a_1}}{\delta A_{\mu}^{a}\,\delta A_{\nu}^{b}\,}\,\frac{\delta\,v_{\mu_2}^{cl\,a_2} }{\delta A_{\rho}^{c}}\,+\,
\frac{\delta^2 v_{\mu_1}^{cl\,a_1}}{\delta A_{\mu}^{a}\,\delta A_{\rho}^{c}\,}\,\frac{\delta\,v_{\mu_2}^{cl\,a_2} }{\delta A_{\nu}^{b}}\,+\,
\frac{\delta^2 v_{\mu_2}^{cl\,a_2}}{\delta A_{\nu}^{b}\,\delta A_{\rho}^{c}\,}\,\frac{\delta\,v_{\mu_1}^{cl\,a_1} }{\delta A_{\mu}^{a}}\,\Ra\,,\,\mu_{i}=(+,i).
\eeqar
The only non-zero first order derivatives of the classical fields in respect to Reggeon fields are the following ones:
\beq\label{BV4}
\frac{\delta\,v_{+}^{cl\,a_1}(y)}{\delta\,A_{+}^{a}(x)}\,=\,\delta^{a\,a_1}\delta^{2}(x_{\bot}\,-\,y_{\bot})\,\delta(x^{+}\,-y^{+})\,
\eeq
and
\beq\label{BV5}
\frac{\delta\,v_{i}^{cl\,a_1}(y)}{\delta\,A_{-}^{a}(x)}\,=\,\delta^{a\,a_1}\delta^{2}(x_{\bot}\,-\,y_{\bot})\,\tilde{G}^{-\,0}_{y^{-}\,x^{-}}\,\D_{i\,y}\,,
\eeq
see definition of $\tilde{G}^{-\,0}_{x^{-}\,y^{-}}$ in Appendix B and calculations in \cite{Our2,Our3}. The  second order non-zero derivatives of the classical fields with respect to the Reggeon fields we account only to the $g$ order, the first second-order derivative is the following one:
\beq\label{BV6}
\frac{\delta^2\,v_{+}^{cl\,a_1}(z)}{\delta A_{+}^{a}(x)\,\delta A_{-}^{b}(y)}\,=\,
-\,2\,g\,f^{a_1 a b }\,\Box^{-1}_{z z_1}\,\left[\,\delta^{2}(x_{\bot}\,-\,z_{1\,\bot})\,\delta^{2}(y_{\bot}\,-\,z_{1\,\bot})\,\delta(x^{+}\,-z_{1}^{+})
\tilde{G}^{-\,0}_{z_{1}^{-}\,y^{-}}\,\D_{i\,z_1}^{2}\right]\,
\eeq
also see \cite{Our2,Our3}.
The second one has the following form 
\beqar\label{BV7}
&\,&\frac{\delta^2\,v_{i}^{cl\,a_1}(z)}{\delta A_{-}^{a}(x)\,\delta A_{-}^{b}(y)}\,=\,g\,f^{a_1 a b}\, \\
&\,&
\int d^4 z_1 \,\Box^{-1}_{z z_1}\,\delta^{2}(x_{\bot}\,-\,z_{1\,\bot})\,\delta^{2}(y_\bot - z_{1\,\bot})\,
\Le\,
\tilde{G}^{-\,0}_{z_{1}^{-}\,y^{-}}\,\tilde{G}^{-\,0}_{y^{-}\,x^{-}}\,
-\,\tilde{G}^{-\,0}_{z_{1}^{-}\,x^{-}}\,\tilde{G}^{-\,0}_{x^{-}\,y^{-}}\,\Ra\,
\D_{i\,z_1}\,\D_{j\,z_1}^{2}
\eeqar
where
\beq\label{BV8}
\Box^{-1}_{z z_1}\,=\,D_{sc}(z, z_1)\,=\,-\,\int\,\frac{d^{4} k}{(2 \pi)^{4}}\,\frac{e^{- \imath (z -z_1) k}}{k^2}\,,
\eeq
and the last one
\beq\label{BV9}
 \frac{\delta^2\,v_{i}^{cl\,a_1}(z)}{\delta A_{+}^{a}(x)\,\delta A_{-}^{b}(y)}\,=\, \frac{1}{2}\,g\,f^{a_1 a b}\,
\Le \tilde{G}^{+\,0}_{z^{+}\,x^{+}}\,-\,\tilde{G}^{+\,0}_{x^{+}\,z^{+}}\, \Ra\,\delta^{2}(z_{\bot}\,-\,x_{\bot})\,\delta^{2}(z_{\bot}\,-\,y_{\bot})\,
\tilde{G}^{-\,0}_{z^{-}\,y^{-}}\,\D_{i\,y}\,.
\eeq
We note, that there are additional contributions to the second order derivatives of higher perturbative orders which we do not consider here.

 There is the first triple Reggeon vertex we calculate:
\beqar\label{BV10}
&\,&\Le K^{a b c }_{x y z}\Ra_{0\,YM}^{+ + -}\, = \,\int\, d^{4} w\, \Le\,\frac{\delta^{3}\,L_{YM}(v_{i}^{cl}(A),\,v_{+}^{cl}(A))}
{\delta A_{+}^{a}(x)\,\delta A_{+}^{b}(y)\, \delta A_{-}^{c}(z) }\,\Ra_{A\,=\,0}\,=\,\nonumber \\ 
&=&\int\, d^{4} w\,\int d^4 w_1 \,\int d^4 w_2\, 
\frac{\delta^{2}\,L_{YM}}{\delta v_{i}^{cl\,a_1}(w_1)\,\delta v_{+}^{cl\,a_2}(w_2)}\,\Le\,
\frac{\delta^2 v_{i}^{cl\,a_1}(w_1)}{\delta A_{+}^{a}(x)\,\delta A_{-}^{c}(z)\,}\,\frac{\delta\,v_{+}^{cl\,a_2}(w_2) }{\delta A_{+}^{b}(y)}\,+\,\right.\nonumber \\
&+&\,
\left.
\frac{\delta^{2}\,L_{YM}}{\delta v_{i}^{cl\,a_1}(w_1)\,\delta v_{+}^{cl\,a_2}(w_2)}\,
\frac{\delta^2 v_{i}^{cl\,a_1}(w_1)}{\delta A_{+}^{b}(y)\,\delta A_{-}^{c}(z)\,}\,\frac{\delta\,v_{+}^{cl\,a_2}(w_2) }{\delta A_{+}^{a}(x)}\,\Ra
\eeqar 
that gives
\beq\label{BV11}
\Le K^{a b c }_{x y z}\Ra_{0\,YM}^{+ + -}\, =\,g\,f^{a b c }\,
\Le \tilde{G}^{+\,0}_{x^{+}\,y^{+}}\,-\,\tilde{G}^{+\,0}_{y^{+}\,x^{+}} \Ra\, \delta^{2}(z_{\bot}\,-\,x_{\bot})\,\delta^{2}(z_{\bot}\,-\,y_{\bot})\,\D_{i\,z}^{2}\,.
\eeq
The second triple Reggeon vertex of interests reads as:
\beqar\label{BV12}
&\,&\Le K^{a b c }_{x y z}\Ra_{0\,YM}^{+ - -}\, = \,\int\, d^{4} w\, \Le\,\frac{\delta^{3}\,L_{YM}(v_{i}^{cl}(A),\,v_{+}^{cl}(A))}
{\delta A_{+}^{a}(x)\,\delta A_{-}^{b}(y)\, \delta A_{-}^{c}(z) }\,\Ra_{A\,=\,0}\,=\,\nonumber \\ 
&=&
\int\, d^{4} w\,\int d^4 w_1 \,\int d^4 w_2\, 
\frac{\delta^{2}\,L_{YM}}{\delta v_{i}^{cl\,a_1}(w_1)\,\delta v_{+}^{cl\,a_2}(w_2)}\,
\frac{\delta^2 v_{i}^{cl\,a_1}(w_1)}{\delta A_{-}^{b}(y)\,\delta A_{-}^{c}(z)\,}\,\frac{\delta\,v_{+}^{cl\,a_2}(w_2) }{\delta A_{+}^{a}(x)}\,+\,\nonumber \\
&+&
\int\, d^{4} w\,\int d^4 w_1 \,\int d^4 w_2\, 
\frac{\delta^{2}\,L_{YM}}{\delta v_{i}^{cl\,a_1}(w_1)\,\delta v_{+}^{cl\,a_2}(w_2)}\,\Le\,
\frac{\delta^2 v_{+}^{cl\,a_2}(w_2)}{\delta A_{+}^{a}(x)\,\delta A_{-}^{b}(y)\,}\,\frac{\delta\,v_{i}^{cl\,a_1}(w_1) }{\delta A_{-}^{c}(z)}\,+\,\right. \nonumber \\
&+&
\left.
\frac{\delta^2 v_{+}^{cl\,a_2}(w_2)}{\delta A_{+}^{a}(x)\,\delta A_{-}^{c}(z)\,}\,\frac{\delta\,v_{i}^{cl\,a_1}(w_1) }{\delta A_{-}^{b}(y)}\,\Ra\,.
\eeqar
The first term in the r.h.s. of \eq{BV12} provides:
\beqar\label{BV13}
&\,&\int\, d^{4} w\,\int d^4 w_1 \,\int d^4 w_2\, 
\frac{\delta^{2}\,L_{YM}}{\delta v_{i}^{cl\,a_1}(w_1)\,\delta v_{+}^{cl\,a_2}(w_2)}\,
\frac{\delta^2 v_{i}^{cl\,a_1}(w_1)}{\delta A_{-}^{b}(y)\,\delta A_{-}^{c}(z)\,}\,\frac{\delta\,v_{+}^{cl\,a_2}(w_2) }{\delta A_{+}^{a}(x)}\,=\,\nonumber \\
&=&
g f^{a b c} \delta^{2}(x_\bot-z_{\bot})
\int dw^{-}\,d z_{1}^{+}\Le -\D_{i\,x}\D_{i\,y} \Ra
 D_{sc}(w^{-},x^{+},x_{\bot};z^{-},z_{1}^{+},y_{\bot}) \tilde{G}^{-\,0}_{z^{-}\,y^{-}} \D_{j\,x}^{2} -(y \leftrightarrow z)\,.
\eeqar
Taking into account that
\beq\label{BV14}
\Le -\D_{i\,x}\D_{i\,y} \Ra\,\int dw^{-}\,d z_{1}^{+}\,D_{sc}(w^{-},x^{+},x_{\bot};z^{-},z_{1}^{+},y_{\bot})\,=\,-\D_{i\,x}\D_{i\,y} \,D_{0}(x_{\bot}, y_{\bot})\,=\,-\delta^{2}(x_{\bot}\,-\, y_{\bot})
\eeq
see \eq{Pro61} definition,
we obtain finally for this contribution:
\beqar\label{BV1333}
&\,&\int\, d^{4} w\,\int d^4 w_1 \,\int d^4 w_2\, 
\frac{\delta^{2}\,L_{YM}}{\delta v_{i}^{cl\,a_1}(w_1)\,\delta v_{+}^{cl\,a_2}(w_2)}\,
\frac{\delta^2 v_{i}^{cl\,a_1}(w_1)}{\delta A_{-}^{b}(y)\,\delta A_{-}^{c}(z)\,}\,\frac{\delta\,v_{+}^{cl\,a_2}(w_2) }{\delta A_{+}^{a}(x)}\,=\,\nonumber \\
&=&
g\,f^{a b c}\,\delta^{2}(x_\bot\,-\,z_{\bot})\,\delta^{2}(x_{\bot}\,-\, y_{\bot})\,\Le\,
 \tilde{G}^{-\,0}_{y^{-}\,z^{-}}\,-\,\tilde{G}^{-\,0}_{z^{-}\,y^{-}}\Ra\,\D_{i\,x}^{2}\,\,.
\eeqar
The two last terms in the r.h.s. of \eq{BV12} are proportional to
\beq\label{BV1444}
\int d z_{1}^{+}\int  d w^{-}\,\D_{-\,w}\,\Le \Box^{-1}_{w z_1}\,\tilde{G}^{-\,0}_{w^{-}\,z^{-}}\,\Ra\,=\,0\,, 
\eeq
therefore \eq{BV1333} expression is the final answer for the second vertex:
\beq\label{BV15}
\Le K^{a b c }_{x y z}\Ra_{0\,YM}^{+ - -}\, =\,g\,f^{a b c}\,\delta^{2}(x_\bot\,-\,z_{\bot})\,\delta^{2}(x_{\bot}\,-\, y_{\bot})\,\Le\,
 \tilde{G}^{-\,0}_{y^{-}\,z^{-}}\,-\,\tilde{G}^{-\,0}_{z^{-}\,y^{-}}\Ra\,\D_{i\,x}^{2}\,.
\eeq
The additional contributions to the vertices are coming from the Lipatov's effective currents expression in the Lagrangian:
\beq\label{BV16}
S_{curr}\,=\,-\,\frac{1}{N}\int d^4 w\, \,tr\left[ v_{+}^{a_1\,cl}\,O^{a_1\, a_2}(v_{+}^{cl})\,\D_{i}^{2}A_{-}^{a_2}\right]\,
\eeq	
see \cite{Our1,Our2}.
We obtain correspondingly:
\beq\label{BV17}
\Le K^{a b c }_{x y z}\Ra_{0\,curr}^{+ + -}\,=\,-\,\frac{1}{2}\,g\,f^{a b c }\,
\Le \tilde{G}^{+\,0}_{x^{+}\,y^{+}}\,-\,\tilde{G}^{+\,0}_{y^{+}\,x^{+}} \Ra\, \delta^{2}(z_{\bot}\,-\,x_{\bot})\,\delta^{2}(z_{\bot}\,-\,y_{\bot})\,\D_{i\,z}^{2}\,,
\eeq
and
\beq\label{BV18}
\Le K^{a b c }_{x y z}\Ra_{0\,curr}^{+ - -}\, =\,-2\,g\,f^{a b c}\,\delta^{2}(y_\bot-x_{\bot})\,\delta^{2}(z_{\bot}- x_{\bot})\,\Le
\tilde{G}^{-\,0}_{y^{-}\,z^{-}} \,-\,\tilde{G}^{-\,0}_{z^{-}\,y^{-}}\Ra\,\D_{i\,x}^{2}\,,
\eeq
here the \eq{BV1444} identity was used again.

 Finally, summing up all contributions we obtain for the vertices:
\beq\label{BV19}
\Le K^{a b c }_{x y z}\Ra_{0}^{+ + -}\,=\,\frac{1}{2}\,g\,f^{a b c }\,
\Le \tilde{G}^{+\,0}_{x^{+}\,y^{+}}\,-\,\tilde{G}^{+\,0}_{y^{+}\,x^{+}} \Ra\, \delta^{2}(z_{\bot}\,-\,x_{\bot})\,\delta^{2}(z_{\bot}\,-\,y_{\bot})\,\D_{i\,z}^{2}\,,
\eeq
and
\beq\label{BV20}
\Le K^{a b c }_{x y z}\Ra_{0}^{+ - -} =-\,g\,f^{a b c}\,\Le
\tilde{G}^{-\,0}_{y^{-}\,z^{-}} \,-\,\tilde{G}^{-\,0}_{z^{-}\,y^{-}}\Ra\,\delta^{2}(y_\bot-x_{\bot})\,\delta^{2}(z_{\bot}- x_{\bot})\,\D_{i\,x}^{2}\,.
\eeq
The Fourier transform of the vertices, in turn,  provides:
\beqar\label{BV21}
\Le \hat{K}^{a b c }_{p_1 p_2 p_3}\Ra_{0}^{+ + -}\,&=&\,-\,\frac{\imath}{2}\,g\,f^{a b c }\,(2 \pi)^7\,p_{3\,i}^{2}
\Le \frac{1}{p_{1\,+}\,+\,\imath\,\varepsilon}\,+\, \frac{1}{p_{1\,+}\,-\,\imath\,\varepsilon}\Ra\,\nonumber \\
&\,&
\,\delta^{2}(p_{1\,\bot} + p_{2\,\bot} +p_{3\,\bot})\,
\delta(p_{1\,+} + p_{2\,+})\,
\delta(p_{1\,-})\,\delta(p_{2\,-})\,\delta(p_{3\,-})\,\delta(p_{3\,+})\,,
\eeqar
and correspondingly
\beqar\label{BV22}
\Le \hat{K}^{a b c }_{p_1 p_2 p_3}\Ra_{0}^{+ - -}\,&=&\,\imath\,g\,f^{a b c }\,(2 \pi)^7\,p_{1\,i}^{2}\,
\Le \frac{1}{p_{2\,-}\,+\,\imath\,\varepsilon}\,+\, \frac{1}{p_{2\,-}\,-\,\imath\,\varepsilon}\Ra\,\nonumber \\
&\,&
\,\delta^{2}(p_{1\,\bot} + p_{2\,\bot} +p_{3\,\bot})\,
\delta(p_{2\,-} + p_{3\,-})\,
\delta(p_{2\,+})\,\delta(p_{3\,+})\,\delta(p_{1\,-})\,\delta(p_{1\,+})\,.
\eeqar
Here the $\tilde{G}^{\pm\,0}_{x^{\pm}\,y^{\pm}}\,=\,\theta(x^{\pm}\,-\,y^{\pm})$ representation of the $\tilde{G}^{\,0}$ Green's function was used, see Appendix B for the details.

 The interesting property of the obtained \eq{BV19}-\eq{BV20} vertices is their non-symmetry with respect to the numerical coefficients in the front of the expressions.
This difference is a consequences of the gauge chosen for the calculations, we use the $v_{-}\,=\,0\,$ gauge which is a non-covariant one and which destroys the target-projectile symmetry
of the problem when two identical particles is scattering as well as a covariance of the Lagrangian with respect to the gauge transforms. Nevertheless, the full amplitude of the interests 
must restore the invariance and symmetry, it is achieved  by the convolution of the Green's function with the impact factors calculated within the scheme with the same gauge. The simplest example of the 
target-projectile symmetry restoring is presented in Appendix E.

\section{One loop RFT correction to the correlator of two reggeized gluons }

 In paper \cite{Our5} the following equation of the correlator of reggeized gluons to leading  RFT order was obtained:
\beqar\label{Corr1}
&\,&\D_{\bot\,x}^{2}< A_{+}^{a}(x^{+},x_{\bot}) A_{-}^{a_1}(y^{-},y_{\bot})>\, = \,-\,\imath\,\delta^{a a_1}\,\delta(x^{+})\,\delta(y^{-})\,\delta^{2}(x_{\bot} - y_{\bot})\,+\,\nonumber \\
&+&
\int d^{2} z_{\bot}\int dz^{+}_{1} d^{2} z_{1\,\bot}\, K^{b_1 b_2 a}_{+ + -}(x^{+},z_{\bot}; z_{1}^{+}, z_{1 \bot}; x_{\bot})\,
< A_{+}^{b_1}(x^{+},z_{\bot}) A_{+}^{b_2}(z_{1}^{+}, z_{1 \bot}) A_{-}^{a_1}(y^{-},y_{\bot}) >\,+\,\nonumber \\
&+&
\int d^{2} z_{\bot}\int dz^{-}_{1} d^{2} z_{1\,\bot}\, K^{b_1 b_2 a}_{+ - -}(z_{\bot}; z_{1}^{-}, z_{1 \bot}; x^{-}, x_{\bot})
< A_{+}^{b_1}(x^{+},z_{\bot}) A_{-}^{b_2}(z_{1}^{-}, z_{1 \bot}) A_{-}^{a_1}(y^{-},y_{\bot}) >.
\eeqar
The equation was truncated in comparison to the expression in \cite{Our5} in order to keep in only vertices of the leading QCD order.
We note immediately, that the second term in the r.h.s. of the equation depends on $x^-$ variable, whereas the correlator in the l.h.s. does not. This discrepancy means a violation
of the kinematical conditions \eq{Ef4}, which in fact were known as valid only when the only first term 
in an expansion of the Reggeon fields with respect to the $x_{\pm}\,\propto\,|t|/s$ small parameter is taken into account,  see \cite{LipatovEff,LipatovEff1,Our4}. Therefore, generalizing the approach, we have to consider
the Regeon fields as four dimensional ones:
\beq\label{Corr101}
A_{+}(x^{+}, x_{\bot})\,\rightarrow\,{\cal B}_{+}(x^{+},x^{-},x_{\bot})\,=\,A_{+}(x^{+}, x_{\bot})\,+\,{\cal D}_{+}(x^{+},x^{-},x_{\bot})\,,\,\,\,\,
{\cal D}_{+}(x^{+},x^{-}=0,x_{\bot})\,=\,0\,
\eeq
and
\beq\label{Corr1011}
A_{-}(x^{-}, x_{\bot})\,\rightarrow\,{\cal B}_{-}(x^{+},x^{-},x_{\bot})\,=\,A_{-}(x^{-}, x_{\bot})\,+\,{\cal D}_{-}(x^{+},x^{-},x_{\bot})\,,\,\,\,\,
{\cal D}_{-}(x^{+}=0,x^{-},x_{\bot})\,=\,0\,.
\eeq
The additional ${\cal D}$ fields represent contributions which usually are neglected in the LLA calculations due their additional kinematical $|t|/s$ suppression, 
as we will see later this is indeed the case. We also note, that we do not consider the contributions from the correlators of two ${\cal D}$
fields, i.e. we do not consider the full correlator of ${\cal B}$ fields. Instead, we limit
the calculations only by "interference" part of the contributions.
Therefore we have for the \eq{Corr1} correlator:
\beqar\label{Corr102}
&\,&< A_{+}^{a}(x^{+},x_{\bot}) A_{-}^{a_1}(y^{-},y_{\bot})>\,\rightarrow\, \\
&\,&
< {\cal B}_{+}^{a}(x^{+},x^{-}\,x_{\bot}) A_{-}^{a_1}(y^{-},y_{\bot})>\,=\,
< A_{+}^{a}(x^{+},x_{\bot}) A_{-}^{a_1}(y^{-},y_{\bot})>\,+\,< {\cal D}_{+}^{a}(x^{+},x^{-},x_{\bot}) A_{-}^{a_1}(y^{-},y_{\bot})>\,\nonumber
\eeqar
that in turn provides:
\beqar\label{Corr103}
&\,&\D_{\bot\,x}^{2}< {\cal D}_{+}^{a}(x^{+},x^{-},x_{\bot}) A_{-}^{a_1}(y^{-},y_{\bot})>\, = \,\\
&=&
\int d^{2} z_{\bot}\int dz^{+}_{1} d^{2} z_{1\,\bot}\, K^{b_1 b_2 a}_{+ + -}(x^{+},z_{\bot}; z_{1}^{+}, z_{1 \bot}; x_{\bot})\,
< A_{+}^{b_1}(x^{+},z_{\bot}) A_{+}^{b_2}(z_{1}^{+}, z_{1 \bot}) A_{-}^{a_1}(y^{-},y_{\bot}) >\,+\nonumber \\
&+&
\int d^{2} z_{\bot}\int dz^{-}_{1} d^{2} z_{1\,\bot}\, K^{b_1 b_2 a}_{+ - -}(z_{\bot}; z_{1}^{-}, z_{1 \bot}; x^{-}, x_{\bot})
< A_{+}^{b_1}(x^{+},z_{\bot}) A_{-}^{b_2}(z_{1}^{-}, z_{1 \bot}) A_{-}^{a_1}(y^{-},y_{\bot}) >\,\nonumber.
\eeqar
This RFT perturbative correction to the two Reggeon fields correlator violates the $x_{\bot}$ dependence of the correlator, some non-trivial dependence on the $x^{-}$ coordinate is arising here, see for comparison \eq{Pro7} expression.
We again underline, that due the absence of the formulation of the approach in terms of 4-d Reggeon fields\footnote{Work in progress.}, we can treat the \eq{Corr103} correlator only perturbatively taking it as the bare expression for the correlator and incerting in the r.h.s. of \eq{Corr103} the expression for the usual Reggeon correlator from \eq{Pro7} .

\begin{figure}[!hb]
\centering
\psfig{file=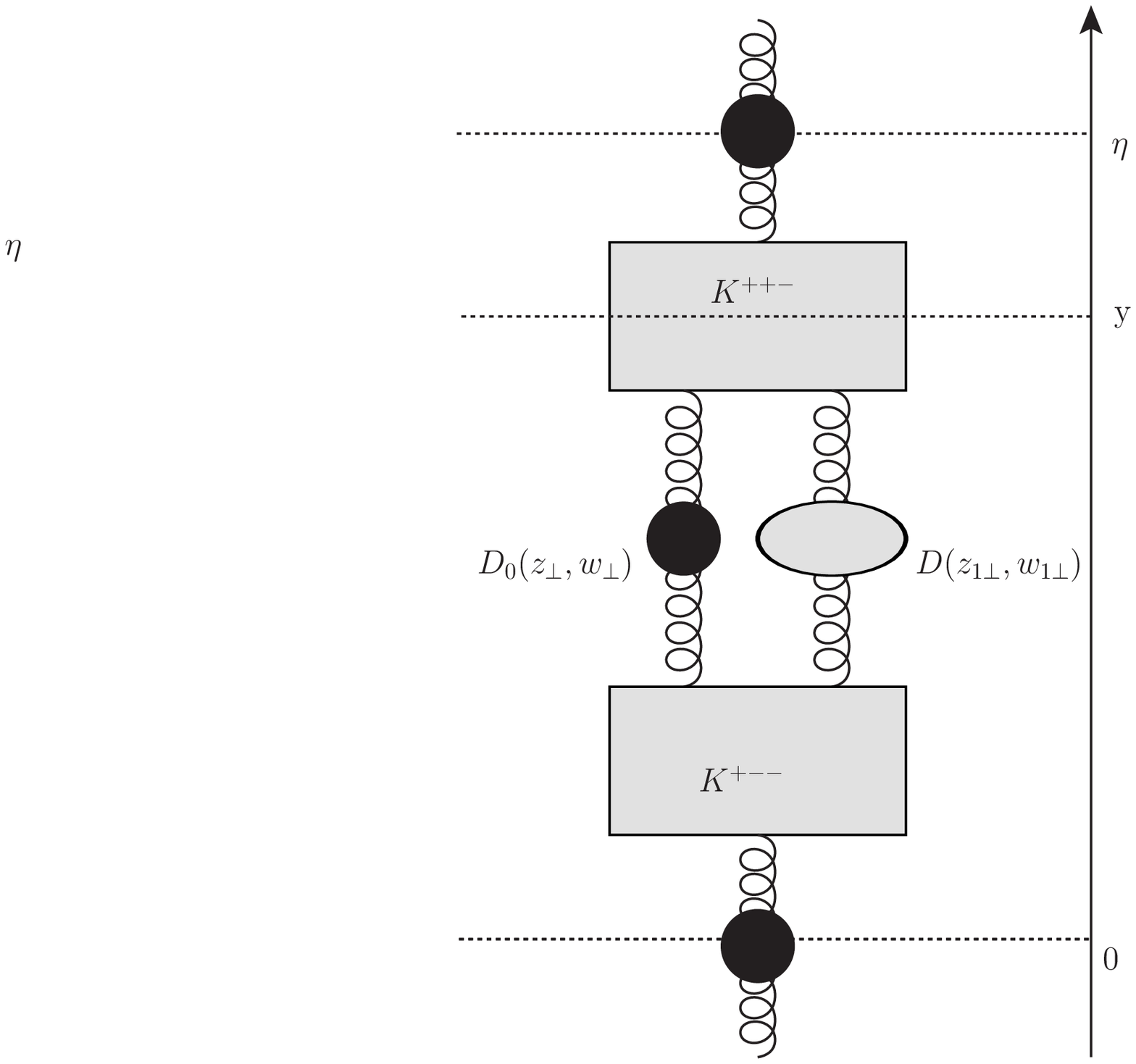,width=90mm} 
\caption{The diagram represents the first term in the r.h.s. of \eq{Corr5} expression.}
\label{Pic1}
\end{figure}

  Now, with the help of \eq{Pro61} function we rewrite the \eq{Corr103} 
equation\footnote{We note, that in comparison with \cite{Our5} equation, the coefficient 2 
is already included in the \eq{BV19}-\eq{BV20} definitions of the vertices.} as following:
\beqar\label{Corr11}
&\,&< {\cal D}_{+}^{a}(x^{+},x^{-},x_{\bot}) A_{-}^{a_1}(y^{-},y_{\bot})>\, = \, \\
&=&
-\int d^{2} z_{\bot}\int dz^{+}_{1} d^{2} z_{1\,\bot}\,\int d^2 z_{2\,\bot} \,D_{0}(x_{\bot},z_{2\,\bot})
K^{b_1 b_2 a}_{+ + -}(x^{+},z_{\bot}; z_{1}^{+}, z_{1 \bot}; z_{2\,\bot})\,
< A_{+}^{b_1} A_{+}^{b_2} A_{-}^{a_1}>\,-\nonumber \\
&-&
\int d^{2} z_{\bot}\int dz^{-}_{1} d^{2} z_{1\,\bot}\int d^2 z_{2\,\bot} D_{0}(x_{\bot},z_{2\,\bot})
K^{b_1 b_2 a}_{+ - -}(z_{\bot}; z_{1}^{-}, z_{1 \bot}; x^{-}, z_{2 \bot})
< A_{+}^{b_1} A_{-}^{b_2} A_{-}^{a_1} >\,. \nonumber
\eeqar
The notation of the vertices are changed here in comparison to \eq{BV19}-\eq{BV20} notations for the shortness.
The expressions for the  triple Reggeon fields correlators   
also were obtained in \cite{Our5} to the leading order precision:
\beqar\label{Corr2}
&\,&\D_{\bot\,x}^{2}< A_{+}^{a}(x^{+},x_{\bot}) A_{+}^{a_1}(y^{+},y_{\bot}) A_{-}^{a_2}(z^{-},z_{\bot}) >\,=\,\\
&=&
\int  d^{2} w_{\bot} d w^{-}_{1} d^{2} w_{1\,\bot}
K^{a_4 a a_3}_{+ - -}(w_{\bot}; x^{-}, x_{\bot}; w^{-}_1, w_{1 \bot})
< A_{+}^{a_4}(x^{+},w_{\bot}) A_{+}^{a_1}(y^{+},y_{\bot}) A_{-}^{a_2}(z^{-},z_{\bot}) A_{-}^{a_3}(w^{-}_{1},w_{1 \bot}) >\,\nonumber
\eeqar
where
\beqar\label{Corr21}
&\,&< A_{+}^{a_4}(x^{+},w_{\bot}) A_{+}^{a_1}(y^{+},y_{\bot}) A_{-}^{a_2}(z^{-},z_{\bot}) A_{-}^{a_3}(w^{-}_{1},w_{1 \bot}) >\,=\, \\
&=&
\imath \delta^{a_4 a_3} \delta(x^+) \delta(w_{1}^{-}) D_{0}(w_{\bot},w_{1 \bot})\,< A_{+}^{a_1} A_{-}^{a_2} >\,+\,
\imath \delta^{a_4 a_2} \delta(x^+) \delta(z^{-}) D_{0}(w_{\bot}, z_{\bot})\,< A_{+}^{a_1} A_{-}^{a_3} >\, \nonumber
\eeqar
see the derivation in \cite{Our5}. Therefore we have for \eq{Corr2}:
\beqar\label{Corr22}
&\,&< A_{+}^{a}(x^{+},x_{\bot}) A_{+}^{a_1}(y^{+},y_{\bot}) A_{-}^{a_2}(z^{-},z_{\bot}) >\,=\,\\
&=&\,-\,\imath\,
\int  d^{2} w_{\bot} \int d w^{-}_{1} d^{2} w_{1\,\bot}\,\int d^{2} w_{2 \bot}\,D_{0}(x_{\bot}, w_{\bot})\,
K^{a_2 a a_3}_{+ - -}(w_{2 \bot}; x^{-}, w_{\bot}; w^{-}_1, w_{1 \bot})\,D_{0}(w_{2 \bot}, z_{\bot})\, \nonumber \\
&\,&
<  A_{+}^{a_1}(y^{+},y_{\bot})  A_{-}^{a_3}(w^{-}_{1},w_{1 \bot}) >\,\delta(x^+)\, \delta(z^{-})\nonumber\,,
\eeqar
we see that there is also dependence on $x^-$ variable in the r.h.s. of \eq{Corr21}, that explains the presence of two terms in the \eq{Corr103} expression.
Correspondingly, we obtain for the second triple Reggeon correlator of interests:
\beqar\label{Corr3}
&\,&\D_{\bot\,x}^{2}< A_{+}^{a}(x^{+},x_{\bot}) A_{-}^{a_1}(y^{-},y_{\bot}) A_{-}^{a_2}(z^{-},z_{\bot}) >\,=\,\\
&=&
\int  d^{2} w_{\bot} d w^{-}_{1} d^{2} w_{1\,\bot}
K^{a_3 a_4 a}_{+ + -}(x^{+}, w_{\bot}; w_{1}^{+}, w_{1 \bot};  x_{\bot})
< A_{+}^{a_3}(x^{+},w_{\bot}) A_{+}^{a_4}(w_{1}^{+},w_{1 \bot}) A_{-}^{a_1}(y^{-},y_{\bot}) A_{-}^{a_2}(z^{-},z_{\bot})  >\,\nonumber
\eeqar
that with the help of \eq{Corr21} gives:
\beqar\label{Corr4}
&\,&< A_{+}^{a}(x^{+},x_{\bot}) A_{-}^{a_1}(y^{-},y_{\bot}) A_{-}^{a_2}(z^{-},z_{\bot}) >\,=\,\\
&=&\,-\, \imath\,
\int  d^{2} w_{\bot} \int d w^{+}_{1} d^{2} w_{1\bot} \int d^{2} w_{2 \bot}\,D_{0}(x_{\bot}, w_{\bot})\,
K^{a_1 a_4 a}_{+ + -}(x^{+}, w_{2 \bot}; w_{1}^{+}, w_{1 \bot};  w_{\bot}) \,D_{0}(w_{2 \bot}, y_{\bot})\,\nonumber \\
&\,&
< A_{+}^{a_4}(w_{1}^{+},w_{1 \bot}) A_{-}^{a_2}(z^{-},z_{\bot}) > \delta(x^+) \delta(y^-) - \nonumber \\
&-&
\imath\,
\int  d^{2} w_{\bot} \int d w^{+}_{1} d^{2} w_{1\bot} \int d^{2} w_{2 \bot}\,D_{0}(x_{\bot}, w_{\bot})\,
K^{a_2 a_4 a}_{+ + -}(x^{+}, w_{2 \bot}; w_{1}^{+}, w_{1 \bot};  w_{\bot}) \,D_{0}(w_{2 \bot}, z_{\bot})\,\nonumber \\
&\,&
< A_{+}^{a_4}(w_{1}^{+},w_{1 \bot}) A_{-}^{a_1}(y^{-},y_{\bot}) > \delta(x^+) \delta(z^-)\,.\nonumber
\eeqar
Taking \eq{Corr11}, \eq{Corr3} and \eq{Corr4} together, we obtain finally equation for the correlator of interests:
\beqar\label{Corr5}
&\,&< {\cal D}_{+}^{a}(x^{+},x^{-},x_{\bot}) A_{-}^{a_1}(y^{-},y_{\bot})>\, = \,\\
&=&
\imath\,\int d^{2} z_{\bot}\, \int d z_{1}^{+} d^{2} z_{1 \bot}\, \int d^{2} z_{2 \bot}\,\int d^{2} w_{\bot}\, \int d w_{1}^{-} d^{2} w_{1 \bot}\,\int d^{2} w_{2 \bot}\,
 \nonumber \\
&\,&
D_{0}(x_{\bot}, z_{2 \bot})\,K^{b_1 b_2 a}_{+ + -}(x^{+}, z_{\bot}; z_{1}^{+}, z_{1 \bot};  z_{2 \bot}) \,
D_{0}(z_{\bot}, w_{\bot})\,K^{a_1 b_1 b_3}_{+ - -}(w_{2 \bot}; x^{-}, w_{\bot}; w^{-}_1, w_{1 \bot})\,D_{0}(w_{2 \bot}, y_{\bot})\,\nonumber \\
&\,&
< A_{+}^{b_2}(z_{1}^{+},z_{1 \bot}) A_{-}^{b_3} (w_{1}^{-},w_{1 \bot} ))>\,\delta(x^{+})\,\delta(y^{-})\,+ \nonumber \\
&+&
\imath\,\int d^{2} z_{\bot}\, \int d z_{1}^{-} d^{2} z_{1 \bot}\, \int d^{2} z_{2 \bot}\,\int d^{2} w_{\bot}\, \int d w_{1}^{+} d^{2} w_{1 \bot}\,\int d^{2} w_{2 \bot}\,
 \nonumber \\
&\,&
D_{0}(x_{\bot}, z_{2 \bot})\,K^{b_1 b_2 a}_{+ - -}(z_{\bot}; z_{1}^{-}, z_{1 \bot}; x^{-} z_{2 \bot}) \,
D_{0}(z_{\bot}, w_{\bot})\,K^{b_2 b_4 b_1}_{+ + -}(x^{+}, w_{2 \bot}; w^{+}_{1}, w_{1 \bot}; w_{\bot})\,D_{0}(w_{2 \bot}, z_{1,\bot})\,\nonumber \\
&\,&
< A_{+}^{b_4}(w_{1}^{+},w_{1 \bot}) A_{-}^{a_1} (y^{-},y_{\bot} ))>\,\delta(x^{+})\,\delta(z^{-}_{1})\,+ \nonumber \\
&+&
\imath\,\int d^{2} z_{\bot}\, \int d z_{1}^{-} d^{2} z_{1 \bot}\, \int d^{2} z_{2 \bot}\,\int d^{2} w_{\bot}\, \int d w_{1}^{+} d^{2} w_{1 \bot}\,\int d^{2} w_{2 \bot}\,
 \nonumber \\
&\,&
D_{0}(x_{\bot}, z_{2 \bot})\,K^{b_1 b_2 a}_{+ - -}(z_{\bot}; z_{1}^{-}, z_{1 \bot}; x^{-} z_{2 \bot}) \,
D_{0}(z_{\bot}, w_{\bot})\,K^{a_1 b_4 b_1}_{+ + -}(x^{+}, w_{2 \bot}; w^{+}_{1}, w_{1 \bot}; w_{\bot})\,D_{0}(w_{2 \bot}, y_{\bot})\,\nonumber \\
&\,&
< A_{+}^{b_4}(w_{1}^{+},w_{1 \bot}) A_{-}^{b_2} (z_{1}^{-},z_{1 \bot} ))>\,\delta(x^{+})\,\delta(y^{-})\,,
\eeqar
see \fig{Pic1}
Taking the integrals in these three terms we obtain:
\beqar\label{Corr6}
&\,&< {\cal D}_{+}^{a}(x^{+},x^{-},x_{\bot}) A_{-}^{a_1}(y^{-},y_{\bot})>\, = \,\\
&=&\,-
\frac{\imath}{2}\,g^{2}\,f^{a b_1 b_2}\,f^{a_1 b_3 b_1}\,\int d z^{+}\,\Le \tilde{G}^{+\,0}_{x^{+}\,z^{+}}\,-\,\tilde{G}^{+\,0}_{z^{+}\,x^{+}} \Ra\,
\int d w^{-} \,\Le \tilde{G}^{-\,0}_{x^{-}\,w^{-}}\,-\,\tilde{G}^{-\,0}_{w^{-}\,x^{-}} \Ra\,D_{0}(x_{\bot}, y_{\bot})\, \nonumber \\
&\,&
< A_{+}^{b_2}(z^{+},x_{\bot}) A_{-}^{b_3}(w^{-},y_{\bot} ))>\,\delta(x^{+})\,\delta(y^{-})\, -\,\nonumber \\
&-&\,
\frac{\imath}{2}\,g^{2}\,f^{a b_1 b_2}\,f^{b_4 b_2 b_1}\,\Le \tilde{G}^{-\,0}_{x^{-}\,0}\,-\,\tilde{G}^{-\,0}_{0\,x^{-}} \Ra\,
\int d w^{+}\,\Le \tilde{G}^{+\,0}_{x^{+}\,w^{+}}\,-\,\tilde{G}^{+\,0}_{w^{+}\,x^{+}} \Ra\, \int d^{2} w_{\bot}\,\int d^2 z_{\bot} \,\nonumber \\
&\,&
D_{0}(x_{\bot}, z_{\bot})\,\Le \D_{i z}^{2}\,\D_{i w}^{2}\,D_{0}(z_{\bot}, w_{\bot})\Ra\,D_{0}(w_{\bot}, z_{\bot})\,
< A_{+}^{b_4}(w^{+},w_{\bot}) A_{-}^{a_1}(y^{-},y_{\bot} ))>\,\delta(x^{+})\,-\,\nonumber \\
&-&\,
\frac{\imath}{2}\,g^{2}\,f^{a b_1 b_2}\,f^{a_1 b_4 b_1}\,\int d z^{-} \Le \tilde{G}^{-\,0}_{x^{-}\,z^-}\,-\,\tilde{G}^{-\,0}_{z^-\,x^{-}} \Ra\,
\int d w^{+}\,\Le \tilde{G}^{+\,0}_{x^{+}\,w^{+}}\,-\,\tilde{G}^{+\,0}_{w^{+}\,x^{+}} \Ra\, \int d^{2} w_{\bot}\,\int d^2 z_{\bot} \,\nonumber \\
&\,&
D_{0}(x_{\bot}, z_{\bot})\,\Le \D_{i z}^{2}\,\D_{i w}^{2}\,D_{0}(z_{\bot}, w_{\bot})\Ra\,D_{0}(w_{\bot}, y_{\bot})\,
< A_{+}^{b_4}(w^{+},w_{\bot}) A_{-}^{b_2}(z^{-},z_{\bot} ))>\,\delta(x^{+})\,\delta(y^-)\,\nonumber 
\eeqar
Treating the expression perturbatively and writing the correlators as propagators 
\beq\label{PrR1}
< A_{+}^{a}(x^{+},x_{\bot}) A_{-}^{b}(y^{-},y_{\bot})>\,=\,\imath\,\delta(x^{+})\,\delta(y^-)\,\delta^{a b}\,D(x_{\bot},y_{\bot})\,
\eeq
and
\beq\label{PrR11}
< {\cal D}_{+}^{a}(x^{+},x^{-},x_{\bot}) A_{-}^{a_1}(y^{-},y_{\bot})>=\,\imath\,{\cal G}^{a a_1}_{+}(x^{+},x^{-},x_{\bot}; y^{-},y_{\bot})\,,
\eeq
where $D(x_{\bot},y_{\bot})\,$ propagator is given by \eq{Pro7}-\eq{Pro1111} expression, see the derivation of \eq{Pro62} in \cite{Our5}, 
we obtain for \eq{Corr6} expression:
\beqar\label{PrR2}
&\,&{\cal G}^{a a_1}_{+}(x^{+},x^{-},x_{\bot}; y^{-},y_{\bot})\,=\, \\
&=&
-\frac{\imath}{2}\,g^2\,N\,\delta^{a a_1}\Le \tilde{G}^{+\,0}_{x^{+}\,0}\,-\,\tilde{G}^{+\,0}_{0\,x^{+}} \Ra\, 
\Le \tilde{G}^{-\,0}_{x^{-}\,0}\,-\,\tilde{G}^{-\,0}_{0\,x^{-}} \Ra\,\Le\,D_{0}(x_{\bot}, y_{\bot})\,D(x_{\bot},y_{\bot})\,-\,\right.\nonumber \\
&-& \left. 
\int d^{2} w_{\bot}\,\int d^2 z_{\bot} \,
D_{0}(x_{\bot}, z_{\bot})\,\Le \D_{i z}^{2}\,\D_{i w}^{2}\,D_{0}(z_{\bot}, w_{\bot})\Ra\,D_{0}(w_{\bot}, z_{\bot})\,D(w_{\bot},y_{\bot})\,-\,\right. \nonumber \\
&-& \left. \,
\int d^{2} w_{\bot}\,\int d^2 z_{\bot} \,
D_{0}(x_{\bot}, z_{\bot})\,\Le \D_{i z}^{2}\,\D_{i w}^{2}\,D_{0}(z_{\bot}, w_{\bot})\Ra\,D_{0}(w_{\bot}, y_{\bot})\,D(w_{\bot},z_{\bot})\,\Ra\,\delta(x^{+})\,\delta(y^-) \nonumber \,.
\eeqar
Performing Fourier transform of the propagator with respect to $x^{+}$ and $x_{\bot}-y_{\bot}$ variables we obtain:
\beqar\label{Corr666}
&\,&\tilde{{\cal G}}^{a a_1}_{+}(p_{+},p_{\bot}; x^{-}, y^{-}; \eta)\,=\, g^2\,N\,\delta^{a a_1}\,\Le \tilde{G}^{-\,0}_{x^{-}\,0}\,-\,\tilde{G}^{-\,0}_{0\,x^{-}} \Ra\, \delta(p_{+})\,\delta(y^-)\nonumber\\ 
&\,&\,\int \frac{dk_{+}}{k_{+}}\,
\int \frac{d^{2} k_{\bot}}{(2\pi)^{2}}\,\Le\,\frac{e^{\,(\eta-y)\,\epsilon(k_{\bot}^{2})}}{k_{\bot}^{2}\,(p_{\bot}\,-\,k_{\bot})^{2}}\,-\,
\frac{e^{\,(\eta-y)\,\epsilon(p_{\bot}^{2})}\,k_{\bot}^{2}}{p_{\bot}^{4}\,(p_{\bot}\,-\,k_{\bot})^{2}}\,-\,
\frac{e^{\,(\eta-y)\,\epsilon(k_{\bot}^{2})}\,(p_{\bot}-k_{\bot})^{2}}{ p_{\bot}^{4}\,k_{\bot}^{2}}\,\Ra\,.
\eeqar
Here a rapidity interval $\eta$ appears as an analog
of an ultraviolet cut-off in the relative longitudinal momenta integration
in the integral on $k^+$ variable, where the rapidity variable $y\,=\,\frac{1}{2}\ln (\Lambda k_{+})\,$ can be  introduced. At the end of the integration it's limit is taking to $Y\,=\,\ln (s/s_{0})$, therefore we have:
\beqar\label{Corr7}
&\,&\tilde{{\cal G}}^{a a_1}_{+}(p_{+},p_{\bot}; x^{-}, y^{-}; Y)\,=\, 
\frac{g^2\,N}{4 \pi^3}\,\delta^{a a_1}\,\Le \tilde{G}^{-\,0}_{x^{-}\,0}\,-\,\tilde{G}^{-\,0}_{0\,x^{-}} \Ra\,\delta(p_{+})\,\delta(y^-)\\ 
&\,&\,
\frac{1}{p_{\bot}^{4}\,}\int d^{2} k_{\bot}\Le \frac{p_{\bot}^{4}\Le e^{\,\epsilon(k_{\bot}^{2})Y\,}\,-\,1\Ra}{k_{\bot}^{2}\,(p_{\bot}\,-\,k_{\bot})^{2}\,\epsilon(k_{\bot}^{2})}\,-\,
\frac{k_{\bot}^{2}\Le e^{ \epsilon(p_{\bot}^{2})Y\,}\,-\,1\Ra}{(p_{\bot}\,-\,k_{\bot})^{2}\,\epsilon(p_{\bot}^{2})}\,-\,
\frac{(p_{\bot}-k_{\bot})^{2}\Le e^{ \epsilon(k_{\bot}^{2})Y\,}\,\,-\,1\Ra}{k_{\bot}^{2}\,\epsilon(k_{\bot}^{2})}\,\Ra\,. \nonumber
\eeqar
 To the leading order this expression reads as:
 \beqar\label{Corr8}
&\,&\tilde{{\cal G}}^{a a_1}_{+}(p_{+},p_{\bot}; x^{-}, y^{-}; Y)\,=\, 
\frac{g^2\,N}{4 \pi^3}\,\ln (s/s_{0})\,\delta^{a a_1}\,\Le \tilde{G}^{-\,0}_{x^{-}\,0}\,-\,\tilde{G}^{-\,0}_{0\,x^{-}} \Ra\,\delta(p_{+})\,\delta(y^-) \\ 
&\,&\,
\frac{1}{p_{\bot}^{4}\,}\int d^{2} k_{\bot}\Le \frac{p_{\bot}^{4}}{k_{\bot}^{2}\,(p_{\bot}\,-\,k_{\bot})^{2}\,}\,-\,
\frac{k_{\bot}^{2}}{(p_{\bot}\,-\,k_{\bot})^{2}\,}\,-\,
\frac{(p_{\bot}-k_{\bot})^{2}}{k_{\bot}^{2}\,}\,\Ra\, \nonumber
\eeqar
and taking into account that the last two integrals are zero due to the 't Hooft-Veltman conjecture in the dimensional regularization scheme, 
see \cite{Leibb}, we correspondingly obtain:
\beq\label{Corr9}
\tilde{{\cal G}}^{a a_1}_{+}(p_{+},p_{\bot}; x^{-}, y^{-}; Y)\,=\,-\,\frac{4}{p_{\bot}^{2}}\,\epsilon(p_{\bot}^{2})\, 
\ln (s/s_{0})\,\delta^{a a_1}\,\Le \tilde{G}^{-\,0}_{x^{-}\,0}\,-\,\tilde{G}^{-\,0}_{0\,x^{-}} \Ra\,\delta(p_{+})\,\delta(y^-)\,,
\eeq
see \eq{Pro9} definition. 
In Appendix D the calculations to the all perturbative orders of \eq{Corr7} are performed and we obtain for the full answer:
\beq\label{Corr10}
\tilde{{\cal G}}^{a a_1}_{+}(p_{+},p_{\bot}; x^{-}, y^{-}; Y)=
\frac{2\delta^{a a_1}}{p_{\bot}^{2}}\Le 1-e^{\epsilon(p_{\bot}^{2}) Y}-\int_{0}^{-\epsilon(p_{\bot}^{2})Y}\frac{d y}{y}\Le e^{-y}-1\Ra \Ra
\Le \tilde{G}^{-0}_{x^{-} 0}-\tilde{G}^{-0}_{0 x^{-}} \Ra \delta(p_{+}) \delta(y^-)\,.
\eeq
We notice, that this correlator arises due the presence in the effective action kinetic term of the  $B_{+}\D_{\bot}^{2} A_{-}$ form with the \eq{Corr101} field's change
applied in some full 4-d effective action. Taking into account also the presences
of the $A_{+}\D_{\bot}^{2} B_{-}$ kinetic term in the same action, we have an additional correction to the two fields correlator which can be obtained from the similar hierarchy
after the variation with respect to $A_{+}$ field\footnote{The considered in the paper Dyson-Schwinger hierarchy is derived from the variation of the effective action with respect to the $A_{-}$ field only.}.  This additional correction can be directly obtained from \eq{Corr10} answer by change of the $+$ sign in the expression 
on the $-$ sign everywhere and corresponding change of the coordinates:
\beq\label{Corr1111}
\tilde{{\cal G}}^{a a_1}_{-}(p_{-},p_{\bot}; x^{+}, y^{+}; Y)=
\frac{2\delta^{a a_1}}{p_{\bot}^{2}}\Le 1-e^{\epsilon(p_{\bot}^{2})Y}-\int_{0}^{-\,\epsilon(p_{\bot}^{2})Y}\frac{d y}{y}\Le e^{-y}-1\Ra \Ra 
\Le \tilde{G}^{-0}_{y^{+}0} -\tilde{G}^{-0}_{0 y^{+}} \Ra \delta(p_{-}) \delta(x^+).
\eeq
The last step in the Fourier transform is the transforms of the theta functions of $x^{-}$ and $y^{-}$ coordinates in \eq{Corr10} and $x^{+}$ and $y^{+}$
coordinates in \eq{Corr1111} correspondingly. We note that the \eq{Corr10}
expression,
for example, is zero at $x^- = 0$, as it must be due the \eq{Corr101} condition, see correspondingly \eq{Corr1011}. Therefore, in order to preserve the property of the expression, we have to regularize 
the Fourier transform of the difference of the theta functions in this limit, we do so with the help of Sokhotski expressions:
\beq\label{Corr311}
\frac{1}{p_{-}\,+\,\imath\,\varepsilon}\,+\, \frac{1}{p_{-}\,-\,\imath\,\varepsilon}\,=\,2\,\mathscr{P}(\frac{1}{p_{-}})
\eeq  
where
\beq\label{Corr312}
\Le \mathscr{P}(\frac{1}{p_{-}}), f(p_-) \Ra\,=\,PV\,\int \,\frac{f(p_{-})}{p_{-}}\,dp_{-}\,=\,\int \,\frac{f(p_{-})\,-\,f(0)}{p_{-}}\,dp_{-}\,.
\eeq
Therefore we obtain for \eq{Corr10}:
\beq\label{Corr40}
\tilde{{\cal G}}^{a a_1}_{+}(p_{+},p_{\bot}, p_{-}; Y)=
\frac{4\,\imath\,\delta^{a a_1}}{p_{\bot}^{2}}\Le 1-e^{\epsilon(p_{\bot}^{2}) Y}-\int_{0}^{-\epsilon(p_{\bot}^{2})Y}\frac{d y}{y}\Le e^{-y}-1\Ra \Ra
\Le \tilde{G}^{-0}_{x^{-} 0}-\tilde{G}^{-0}_{0 x^{-}} \Ra \delta(p_{+})\,\mathscr{P}(\frac{1}{p_{-}})\,.
\eeq
and performing corresponding change of the signs for \eq{Corr1111} the final expression for the reggeized gluons propagator acquired the following form in the momentum space:
\beq\label{Corr12}
D^{a b}(p_{+},p_{\bot}, p^{-}; Y)\,=\,D^{a b}(p_{\bot};Y)\,\delta(p_{+})\,\delta(p_{-})\,+\,\imath\,{\cal G}^{a a_1}_{+}(p_{+},p_{\bot}, p_{-};Y)\,+\,\imath\,
{\cal G}^{a a_1}_{-}(p_{+},p_{\bot}, p_{-}; Y)\,,
\eeq
here we separately wrote in front of the last two terms the $\imath$ from \eq{Corr40}.

 Concerning this form of the corrections to the leading order propagator, we note that in the full amplitude, when the Green's function is convoluted with the impact factors, the corrections give non-zero contributions only for non-eikonal form of the  impact factors. Indeed, expanding the impact factor $\Phi$ in respect to transferred momentum $p_{\pm}$
\beq\label{Corr50}
\Phi\,=\,\sum_{n,m\,=\,0}^{\infty}\,C_{n\,m}\,p_{+}^{n}\,p_{-}^{m}
\eeq
we obtain to leading order when the impact factor has an eikonal form:
\beq\label{Corr51}
A_{0}\,\propto\,\Le \mathscr{P}(\frac{1}{p_{\pm}}), 1 \Ra\,=\,0\,.
\eeq
To the next to leading order we obtain for the corrections in the multi-Regge-kinematics:
\beq\label{Corr52}
A_{1}\,\propto\,\Le \mathscr{P}(\frac{1}{p_{\pm}}), p_{\pm} \Ra\,=\,\int d \, p_{\pm}\,\propto\,1 / \sqrt{s}\,
\eeq
that provides additional suppression of these contributions in the full amplitude. The structure of these contributions of  ${\cal D}_{\pm}$ fields was already mentioned in \cite{Our4} paper,
next orders of the non-eikonal contributions will be suppressed correspondingly as $s^{-n/2}$. We conclude, that these corrections are non-eikonal, i.e. they are beyond the usual BFKL LLA approximation,
and that are suppressed in the final amplitudes by the additional non-logarithmic $1/(\sqrt{s})^{n}$ factor. Indeed, comparing the 
obtained result with a perturbative NNLO contribution of the $\alpha^2\,\ln^2(s)$ order we see that the non-eikonal corrections are only important when the condition
\beq\label{AddCor}
\sqrt{s}\,\ln (s)\,<\,1/\alpha
\eeq  
is satisfied.

\section{Non-linear unitary correction to the propagator of reggeized gluons}

\begin{figure}[!hb]
\centering
\psfig{file=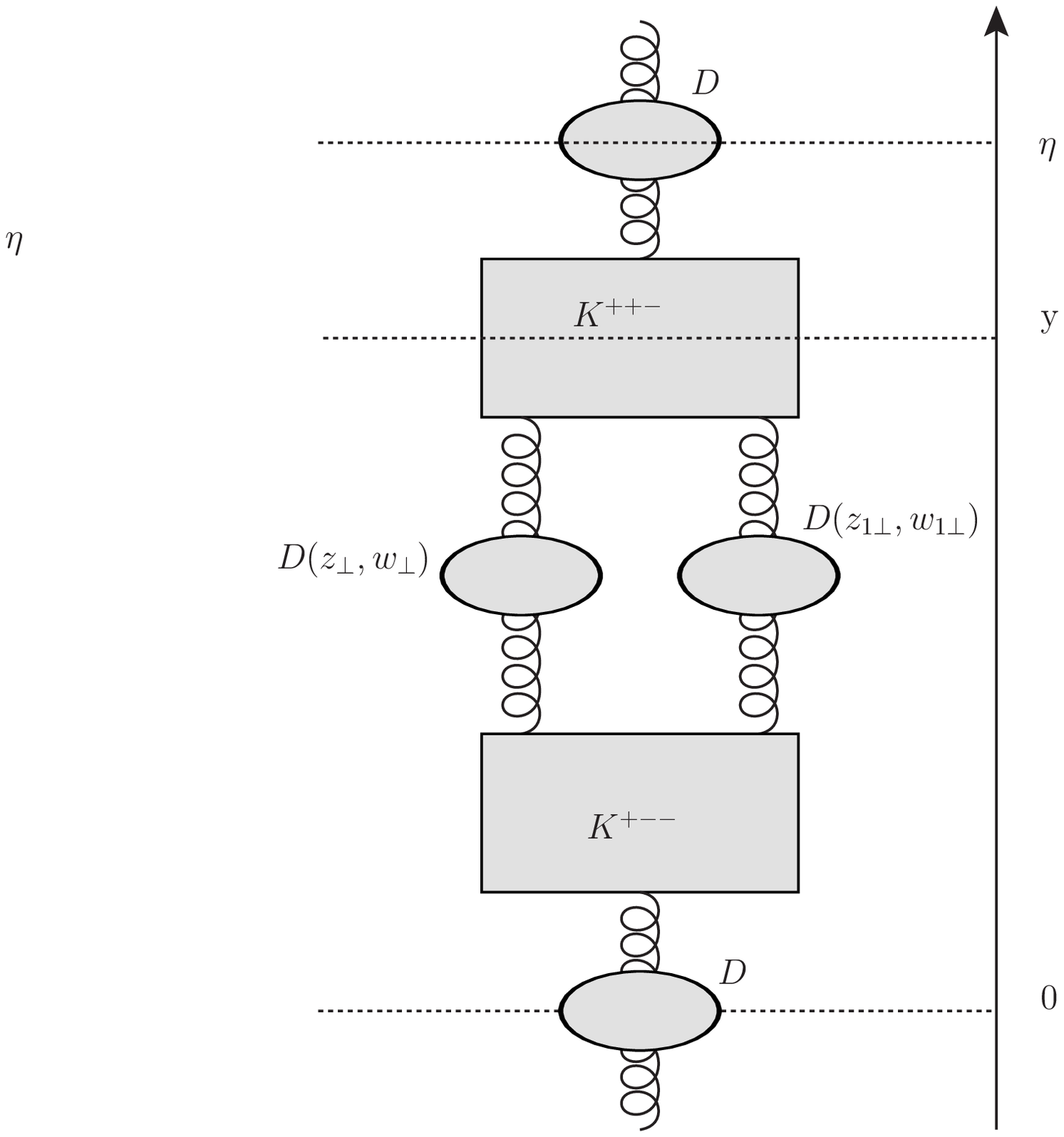,width=90mm} 
\caption{The diagram represents \eq{NL6} expression.}
\label{Pic2}
\end{figure}

 In the above equations for the three and four Reggeon fields correlators, \eq{Corr2} - \eq{Corr3}, we took into account only the bare QCD RFT corrections in the corresponding equations.
The additional one-loop QCD correction in the equations will be provided by \eq{Pro1} vertex, namely the equation for the triple fields correlator with this correction reads as:
\beqar\label{NL1}
&\,&\D_{\bot\,x}^{2}< A_{+}^{a}(x^{+},x_{\bot}) A_{+}^{a_1}(y^{+},y_{\bot}) A_{-}^{a_2}(z^{-},z_{\bot}) >\,=\,\\
&=&
\int  d^{2} w_{\bot} d w^{-}_{1} d^{2} w_{1\,\bot}
K^{a_4 a a_3}_{+ - -}(w_{\bot}; x^{-}, x_{\bot}; w^{-}_1, w_{1 \bot})
< A_{+}^{a_4}(x^{+},w_{\bot}) A_{+}^{a_1}(y^{+},y_{\bot}) A_{-}^{a_2}(z^{-},z_{\bot}) A_{-}^{a_3}(w^{-}_{1},w_{1 \bot}) >\,+\,\nonumber \\
&+&\int  d^{2} w_{\bot}\, K^{a_3 a}_{+ -}(w_{\bot}, x_{\bot})\,
< A_{+}^{a}(x^{+},w_{\bot}) A_{+}^{a_1}(y^{+},y_{\bot}) A_{-}^{a_2}(z^{-},z_{\bot}) >\,,
\eeqar
see \cite{Our5}. Formally, this contribution means the following replacement of the operator in the l.h.s. of the equation:
\beq\label{NL2}
\D_{\bot}^{2}\,\rightarrow\,\D_{\bot}^{2}\,-\,K_{+ -}\,.
\eeq
We note that the propagator \eq{Pro7} is a Green's function of the operator:
\beq\label{NL3}
\Le \D_{\bot}^{2}\delta^{a b}\,-\,K_{+ -}^{a b}\Ra\,D^{b c}\,=\,-\delta^{a c}\,,
\eeq 
see \eq{Pro6} expression. Therefore, account of this type of the one-loop QCD corrections means the replacement
\beq\label{NL4}
D_{0}^{a b}(x_{\bot}, y_{\bot})\,\rightarrow\,D^{a b}(x_{\bot}, y_{\bot})
\eeq
everywhere in the above expressions for the correlators of three and four Reggeon fields. For example, we will obtain instead \eq{Corr21} the following expression
\beqar\label{NL5}
&\,&< A_{+}^{a_4}(x^{+},w_{\bot}) A_{+}^{a_1}(y^{+},y_{\bot}) A_{-}^{a_2}(z^{-},z_{\bot}) A_{-}^{a_3}(w^{-}_{1},w_{1 \bot}) >\,=\, \\
&=&
\imath  \delta(x^+) \delta(w_{1}^{-}) D^{a_4 a_3}(w_{\bot},w_{1 \bot})\,< A_{+}^{a_1} A_{-}^{a_2} >\,+\,
\imath \delta(x^+) \delta(z^{-}) D^{a_4 a_2}(w_{\bot}, z_{\bot})\,< A_{+}^{a_1} A_{-}^{a_3} >\,, \nonumber
\eeqar
which now is fully symmetrical with respect to the color indexes. Using this prescription,  we obtain for \eq{Corr5} correlator:
\beqar\label{NL6}
&\,&< {\cal D}_{+}^{a}(x^{+},x^{-},x_{\bot}) A_{-}^{a_1}(y^{-},y_{\bot})>\, = \,\\
&=&
\imath\,\int d^{2} z_{\bot}\, \int d z_{1}^{+} d^{2} z_{1 \bot}\, \int d^{2} z_{2 \bot}\,\int d^{2} w_{\bot}\, \int d w_{1}^{-} d^{2} w_{1 \bot}\,\int d^{2} w_{2 \bot}\,
 \nonumber \\
&\,&
D(x_{\bot}, z_{2 \bot})\,K^{b_1 b_2 a}_{+ + -}(x^{+}, z_{\bot}; z_{1}^{+}, z_{1 \bot};  z_{2 \bot}) \,
D(z_{\bot}, w_{\bot})\,K^{a_1 b_1 b_3}_{+ - -}(w_{2 \bot}; x^{-}, w_{\bot}; w^{-}_1, w_{1 \bot})\,D(w_{2 \bot}, y_{\bot})\,\nonumber \\
&\,&
< A_{+}^{b_2}(z_{1}^{+},z_{1 \bot}) A_{-}^{b_3} (w_{1}^{-},w_{1 \bot} ))>\,\delta(x^{+})\,\delta(y^{-})\,,
\eeqar
here the one loop QCD correction is accounted for the two fields correlator as well, see \fig{Pic2}.
The transverse structure of the above integral is not changing in comparison to the obtained answer, the only change is in the rapidity dependence of the integral, namely we have instead \eq{Corr666}:
\beq\label{NL7}
\tilde{{\cal G}}^{a a_1}_{+}\,=\, g^2\,N\,\delta^{a a_1}\,\Le \tilde{G}^{-\,0}_{x^{-}\,0}\,-\,\tilde{G}^{-\,0}_{0\,x^{-}} \Ra\,
\int \frac{d^{2} k_{\bot}}{(2\pi)^{2}}\,\frac{1}{k_{\bot}^{2}\,(p_{\bot}\,-\,k_{\bot})^{2}}
\int_{0}^{\eta}\frac{dy}{\pi}\,e^{(2\,\eta-y)\,\epsilon(k_{\bot}^{2})}\,\delta(y^{-})\,\delta(p_{\,+})
\eeq
that gives after the rapidity integration:
\beq\label{NL8}
\tilde{{\cal G}}^{a a_1}_{+}(p_{+},p_{\bot};x^{-}, y^{-}; Y)\,=\, 
\frac{g^2\,N}{4 \pi^3}\,\delta^{a a_1}\,\Le \tilde{G}^{-\,0}_{x^{-}\,0}\,-\,\tilde{G}^{-\,0}_{0\,x^{-}} \Ra\,
\int d^{2} k_{\bot}\,\frac{e^{2\,\epsilon(k_{\bot}^{2})Y\,}\,-\,e^{\,\epsilon(k_{\bot}^{2})Y\,}}{k_{\bot}^{2}\,(p_{\bot}\,-\,k_{\bot})^{2}\,\epsilon(k_{\bot}^{2})}\,\delta(y^{-})\,\delta(p_{+})\,.
\eeq
Formally the first term in the integral is different from the \eq{Corr7} answer by $2$ in the front of the trajectory, therefore, using results of Appendix D, we obtain after the integration on momenta:
\beq\label{NL9}
\tilde{{\cal G}}^{a a_1}_{+}\,=\,
\frac{2\,\delta^{a a_1}}{p_{\bot}^{2}}\,\Le e^{\,\epsilon(p_{\bot}^{2})\,Y}\,-\,e^{2\,\epsilon(p_{\bot}^{2})\,Y}\,-\,
\int_{-\,\epsilon(p_{\bot}^{2})\,Y}^{-\,2\,\epsilon(p_{\bot}^{2})\,Y}\,\frac{d y}{y}\Le e^{-y}\,-\,1\,\Ra \Ra\,
\Le \tilde{G}^{-\,0}_{x^{-}\,0}\,-\,\tilde{G}^{-\,0}_{0\,x^{-}} \Ra\,\delta(y^{-})\,\delta(p_{\,+})\,
\eeq
with corresponding change of \eq{Corr1111} expression and full answer for the propagator \eq{Corr12}. The Fourier transform corresponding to $x^{-}, y^{-}$ coordinates in \eq{NL9} 
must be performed in the same way as it was done in the previous section. We note, that at the large values of rapidity, the non-eikonal suppression of the contributions is convoluted
with the square of the propagator of the reggeized gluons in the expressions, see \fig{Pic2}. In this extend, there are additional non-linear corrections to the one-loop RFT correlator
even in comparison to \eq{Corr12} expression, due their contribution beyond the LLA we can consider them as a kind of the unitarity corrections as well which possibly restore the unitarity
of the S-matrix in the broader kinematic region of the scattering.

\section{Conclusion}

 In the Lipatov's effective action formalism we have an additional source of perturbative and unitary corrections to high energy QCD amplitudes which is based
on the diagrams constructed entirely in terms of Reggeon fields.
In this paper we calculated to the first time the one loop RFT contribution to the propagator of reggeized gluons in the framework of Lipatov's effective action
and Dyson-Schwinger hierarchy of the equations for the correlators of reggeized
gluon fields.
This correction to the propagator is the first and the simplest non-linear RFT correction which combines the effective vertices of the theory to one loop QCD
precision and one RFT loop constructed from the propagators of reggeized gluons.

 The three Reggeon vertices of interests, requested for the calculation of the non-linear corrections, age given by \eq{BV19} - \eq{BV22} to the bare QCD precision. The important properties of the 
obtained expressions are that the pole structure of the vertices is determined by the Green's functions operators in the expression
and can be different depending on the these operator's representation, see Appendix B.
The knowledge of these vertices, in turn, allows to consider the theory with \eq{Sec1} effective action as an usual quantum field theory with the interaction of three Reggeon fields included and correspondingly allows to calculate RFT
perturbative corrections to any object of interests in the theory. The next step in the further development of the theory is the calculation of these vertices to one loop QCD  precision
similarly to done for the vertex of two Reggeon interactions\footnote{Work in progress} in Section 3. The one loop QCD triple vertices  will change the rapidity 
structure of the two reggeized gluons correlator as well and will allow understand better the quantum structure of the theory.

 The main results of the article, therefore, are the corrections to the propagator of reggeized gluons given by contributions in \eq{Corr10}-\eq{Corr12} and \eq{NL9}.
We constructed the one loop RFT corrections to the \eq{Pro1111}
propagator taking into account the  Reggeon fields correlator to one QCD loop precision and bare triple Reggeon vertices. 
The form of the obtained one-loop contribution, \eq{Corr1}, demonstrated the inconsistency of the RFT based on the Reggeon fields of the \eq{Ef4} types only.
This result was expected, see notes in \cite{LipatovEff,Our4}, but surprisingly this modification of the Reggeon field was required already at one RFT loop contribution. 
In general, therefore, the consistent construction of QCD RFT requires an use of the 4-d Reggeon fields, see \eq{Corr101}-\eq{Corr1011} expressions. Correspondingly, obtained corrections
\eq{Corr1111} and \eq{NL9} can  be considered as some bare contributions in closed equations
for the mutual correlators of $\cal{D}_{\pm}$ and $A_{\pm}$ Reggeon fields, similar to the first term in the r.h.s. of \eq{Pro8} for example, which can be derived in the framework with 
 4-d $\cal{B}_{\pm}$ Reggeon fields included.
This construction of the  Lipatov's effective action in terms $\cal{B}_{\pm}$ Reggeon fields is very interesting 
and important task which we postpone for the future work.

  It turns out that the expressions obtained are depend as well on the longitudinal coordinates, see \eq{Corr12}, that violates the only transverse coordinates dependence of the
\eq{Pro1111} propagator. Moreover, there is additional part of the corrections that breaks  the propagator's reggeization. Namely, there is the dependence on the coupling constant in the expression
which is not "sitting" in the reggeized gluon's trajectory exponent. 
The change of the propagator's form is more drastic if we account the  one loop QCD correction in the expressions for the three and four 
Reggeon correlators. In this case the rapidity dependence of the corrections even is more complicated, see \eq{NL9} expression. Another important future of the both non-linear corrections is that they
contribute to the full amplitude only in the case of non-eikonal impact factors accounted, these contributions are non-eikonal,
see also about non-eikonal corrections in \cite{Nestor} for example, 
and require a knowledge of the impact factors of the scattering process to the non-eikonal precision. In correspondence to \cite{Our4} results, these non-eikonal contributions are also additionally suppressed at leading order as $s^{-1/2}$ 
in comparison to the LL contribution of the propagator of reggeized gluons. In this extend, the corrections obtained are beyond the usual LLA of the BFKL calculus but can be important at the kinematic regime with $\sqrt{s}\,\ln(s)\,<\,1/\alpha$ condition satisfied.

	The next important step to be considered in the future research is the calculation of the BFKL Pomeron on the base of the \eq{Corr12} form of the reggeized gluons propagator. Indeed, the infrared divergence of the
obtained propagator is different from the divergence of the usual trajectory function, \eq{Pro9}, by only the coefficients in the front of each $1/\varepsilon^n$ term in the corresponding expansion. Therefore, the interesting subject of the 
future research is the form of the four Reggeon colorless correlator, BFKL Pomeron, obtained on the base of RFT Dyson-Schwinger hierarchy of the equations for the correlators.  
Namely, it must be checked that the infrared divergences will be absent there as well as in the usual case and also the important question to investigate is about the form and rapidity dependence of this modified Pomeron.

 In conclusion we emphasize, that  the article is considered as an additional step to the developing of the high energy QCD RFT which will help clarify the non-linear RFT  unitary corrections to the 
amplitudes of high energy processes.

\newpage
\section*{Appendix A: Bare gluon propagator in light-cone gauge}
\renewcommand{\theequation}{A.\arabic{equation}}
\setcounter{equation}{0}

 In order to reproduce the expression for the gluon fields bare propagator\footnote{We suppress color and coordinate notations in the definition of the propagators below.} in the light-cone gauge we solve the following system of equations
\beq\label{A1}
M^{\,0\,\mu\,\nu}\,G_{\,0\,\nu\,\rho}\,=\,\delta^{\,\mu}_{\,\rho}
\eeq
with
\beq\label{A101}
g^{\,\mu}_{\,\nu}\, =\,\left(
\begin{array}{cccc}
0 & 1 & 0 & 0 \\
1 & 0 & 0 & 0 \\
0 & 0 & 1 & 0 \\
0 & 0 & 0 & 1
\end{array} \right)\,\,\,\,\,\mu\,,\nu\,=\,(+,\,-,\,\bot)\,,
\eeq
see also \eq{A13} below.
The expression for the $M_{\,0\,\mu\,\nu}$ matrix can obtained from the bare gluon's Lagrangian for the gluon's fluctuations field, in light-cone gauge it has the following form:
\beq\label{A11}
L_{0}\, = \,-\,\frac{1}{2}\,\ep_{i}^{a}\,\delta_{\,a\,b}\Le \delta_{ij}\,\Box\, +\D_{i}\,\D_{j} \Ra\,\ep_{j}^{b}\,+\,
\ep_{+}^{a}\,\D_{-}\,\D_{i}\,\ep_{i}^{a}\,-\,\frac{1}{2}\,\ep_{+}^{a}\,\D_{-}^{2}\,\ep_{+}^{a}\,=\,
-\,\frac{1}{2}\,\ep_{\mu}^{a}\,M_{\,0\,\,\mu\,\nu}\,\ep_{\nu}^{b}\,\delta^{\,a\,b}\,,
\eeq
In the following system of equations 
\beqar\label{A12}	
\,M^{\,i\,+}_{\,0}\,G_{\,0\,+\,j}\,+\,M^{\,i\,k}_{\,0}\,G_{\,0\,k\,j}\,=\,\delta^{\,i}_{\,j}\,\nonumber \\
\,M^{\,+\,i}_{\,0}\,G_{\,0\,i\,+}+M^{\,+\,+}_{\,0}\,G_{\,0\,+\,+}=\delta^{\,+}_{\,+}\,\nonumber \\
\,M^{\,+\,i}_{\,0}\,G_{\,0\,i\,j}\,+\,M^{\,+\,+}_{\,0}\,G_{\,0\,+\,j}\,=\,0\, \nonumber \\
\,M^{\,i\,p}_{\,0}\,G_{\,0\,p\,+}\,+\,M^{\,i\,+}_{\,0}\,G_{\,0\,+\,+}\,=\,0\,,
\eeqar
the last two equations we can consider as definitions of corresponding Green's functions:
\beq\label{A2}	
G_{0\,+\,i}\,=\,-M_{0\,+\,+}^{-1}\,\,M^{\,+\,j}_{\,0}\,G_{\,0\,j\,i}\,,
\eeq	
and
\beq\label{A21}
G_{0\,i\,+}\,=\,-M_{0\,i\,j}^{-1}\,\,M^{\,j\,+}_{\,0}\,G_{\,0\,+\,+}\,.
\eeq	
Here for
\beq\label{A3}
M_{\,0\,\,p\,j}\,=\,\delta_{p\,j}\,\Box\, +\D_{p}\,\D_{j}\,,\,\,\,M_{\,0\,p\,-}\,=\,-\,\D_{p}\,\D_{-}\,,\,\,\,M_{\,0\,\,-\,-}\,=\,\D_{-}^{2}\,
\eeq
we have
\beq\label{A2111}
M^{-1}_{0\,ij}(x,y)\,=\,-\,\int\,\frac{d^4 p}{(2\pi)^{4}}\,\frac{e^{-\imath\,p\,(x\,-\,y)}}{p^2}\,\Le
\delta_{ij}\,-\,\frac{p_{i}\,p_{j}}{2\Le p_{-}\,p_{+}\Ra}\Ra\,
\eeq
and correspondingly
\beq\label{A22}
M^{-1}_{0\,+\,+}(x,y)\,=\,-\,\int\,\frac{d^4 p}{(2\pi)^{4}}\,\frac{e^{-\imath\,p\,(x\,-\,y)}}{p^{2}_{-}}\,.
\eeq
Therefore, for the two remaining Green's functions we obtain:
\beq\label{A2222}
\Le\, M^{+\,+}_{\,0}\,-\,M_{\,0}^{\,+\,i}\,M_{0\,i\,j}^{-1}\,\,M^{\,j\,+}_{\,0}\,\Ra\,G_{\,0\,+\,+}\,=\,\delta^{\,+}_{\,+}\,,
\eeq
and
\beq\label{A23}
\Le\, M^{i\,k}_{\,0}\,-\,M_{\,0}^{\,i\,+}\,M_{0\,+\,+}^{-1}\,\,M^{\,+\,k}_{\,0}\,\Ra\,G_{\,0\,k\,j}\,=\,\delta^{\,i}_{\,j}\,.
\eeq
Performing Fourier transform of the functions, we write the \eq{A23} in the following form:
\beq\label{A24}
-\int\,\frac{d^4 p}{(2\pi)^{4}}\,\Le \delta^{\,i\,k}\,p^{\,2} + p^{\,i}\,p^{\,k}  \Ra\,e^{-\imath\,p\,(x\,-\,y)}\,\tilde{G}_{\,0\,k\,j}(p)\,+\,
\int\,\frac{d^4 p}{(2\pi)^{4}}\,p_{-}^{\,2}\,p^{\,i}\,p^{\,k}\,\frac{e^{-\imath\,p\,(x\,-\,y)}}{p_{-}^{\,2}}\,\tilde{G}_{\,0\,k\,j}(p)\,=\,\delta^{\,i}_{\,j}\,
\eeq
that provides:
\beq\label{A25}
G_{0\,i\,j}(x,y)\,=\,-\,\int\,\frac{d^4 p}{(2\pi)^{4}}\,\frac{e^{-\imath\,p\,(x\,-\,y)}}{p^{\,2}}\,\delta_{\,i\,j}\,.
\eeq
Correspondingly, for \eq{A2222} we have:
\beq\label{A26}
-\int\,\frac{d^4 p}{(2\pi)^{4}}\,p_{\,-}^{\,2}\,e^{-\imath\,p\,(x\,-\,y)}\,\tilde{G}_{\,0\,+\,+}(p)\,+\,
\int\,\frac{d^4 p}{(2\pi)^{4}}\,\frac{p_{\,-}^{\,2}\,p^{\,i}\,p^{\,j}}{p^{\,2}} \Le \delta_{\,i\,j}\,-\,\frac{p_{i}\,p_{j}}{2\Le p_{-}\,p_{+}\Ra} \Ra\,e^{-\imath\,p\,(x\,-\,y)}\,
\tilde{G}_{\,0\,+\,+}(p)\,\,=\,\delta^{\,+}_{\,+}\,,
\eeq
that can be rewritten as
\beq\label{A261}
\int\,\frac{d^4 p}{(2\pi)^{4}}\,e^{-\imath\,p\,(x\,-\,y)}\,\Le
-\,p_{\,-}^{\,2}\,+\,p_{\,-}^{\,2}\,\frac{p^{\,i}\,p^{\,i}}{p^{\,2}}\,-\,p_{\,-}^{\,2}\,\frac{\Le p^{\,i}\,p_{\,i}\Ra\,\Le p^{\,j}\,p_{\,j}\Ra}{2\,p^{\,2} \Le p_{\,+} p_{\,-} \Ra}\,\Ra\,
\tilde{G}_{\,0\,+\,+}(p)\,\,=\,\delta^{\,+}_{\,+}\,.
\eeq
Writing this expression as
\beq\label{A262}
\int\,\frac{d^4 p}{(2\pi)^{4}}\,e^{-\imath\,p\,(x\,-\,y)}\,\Le
-\,p_{\,-}^{\,2}\,-\,p_{\,-}^{\,2}\,\frac{p^{\,i}\,p_{\,i}}{p^{\,2}}\,-\,p_{\,-}^{\,2}\,\frac{\Le p^{\,i}\,p_{\,i}\Ra\,\Le p^{\,j}\,p_{\,j}\Ra}{2\,p^{\,2} \Le p_{\,+} p_{\,-} \Ra}\,\Ra\,
\tilde{G}_{\,0\,+\,+}(p)\,\,=\,\delta^{\,+}_{\,+}\,.
\eeq
we obtain finally for the Green's function
\beq\label{A263}
G_{0\,+\,+}(x,y)\,=\,-\,\int\,\frac{d^4 p}{(2\pi)^{4}}\,\frac{e^{-\imath\,p\,(x\,-\,y)}}{p^{\,2}}\,\frac{2\,p_{\,+}}{p_{\,-}}\,.
\eeq
Inserting \eq{A25} and \eq{A263} functions in \eq{A2}-\eq{A21} definitions we obtain for the last two Green's functions:
\beq\label{A5}
G_{\,0\,i\,+}\,=\,G_{\,0\,+\,i}\,=\,\int\,\frac{d^4 p}{(2\,\pi)^{4}}\,\frac{e^{-\,\imath\,p\,(x\,-\,y)}}{p^{\,2}}\,\frac{p_{\,i}}{p_{-}}\,.
\eeq	
Now, introducing the following vector in light-cone coordinates
\beq\label{A1101}
n_{\,\mu}^{+}\,=\,(1,\,0,\,0_{\bot})\,,\,\,\,\,\mu\,=\,(+,\,-,\,\bot)
\eeq
 we can write the whole propagator as
\beq\label{A1202}
G_{\,0\,\mu\,\nu}(x,\,y)\,=\,\int\,\frac{d^{4} p}{(2\,\pi)^{4}}\,\frac{e^{-\,\imath\,p\,(x\,-\,y)}}{p^{\,2}}\,\Le\,g_{\mu\,\nu}\,-\,g_{\mu\,\sigma}\,g_{\nu\,\rho}\,
\frac{p^{\,\sigma}\,n^{\,+\,\rho}\,+\,p^{\,\rho}\,n^{\,+\,\sigma}\,}
{p^{\,\rho}\,n_{\,\rho}^{\,+}}\,\Ra\,
\eeq                         
with
\beq\label{A13}
g_{\,\mu\,\nu}\, =\, g^{\,\mu\,\nu}\,= \left(
\begin{array}{cccc}
0 & 1 & 0 & 0 \\
1 & 0 & 0 & 0 \\
0 & 0 & -1 & 0 \\
0 & 0 & 0 & -1
\end{array} \right)\,\,\,\,\,\mu\,,\nu\,=\,(+,\,-,\,\bot)\,
\eeq
and where Kogut-Soper convention, \cite{}, for the light-cone notations and scalar product is used:
\beq\label{A31}
p\,x\,=\,p_{+}\,x^{+}\,+\,p_{-}\,x^{-}\,+\,p_{i}\,x^{i}\,.
\eeq

\newpage
\section*{Appendix B: Lipatov's effective current}
\renewcommand{\theequation}{B.\arabic{equation}}
\setcounter{equation}{0}

For the arbitrary representation of gauge field $v_{+}\,=\,\imath\,T^{a}\,v_{+}^{a}$ with
$D_{+}\,=\,\D_{+}\,-\,g\,v_{+}$, we can consider
the following representation of $O$ and $O^{T}$ operators
\footnote{Due the light cone gauge we consider here only $O(x^{+})$ operators. 
The construction of the representation of the $O(x^{-})$ operators can be done similarly. We also note, that the integration is assumed for 
repeating indexes in expressions below if it is not noted otherwise. }:
\beq\label{B11}
O_{x}\,=\,\delta^{a\, b}\,+\,g\,\int\,d^{4}y\,G_{x y}^{+\,a\, a_{1}}\, \Le v_{+}(y)\Ra_{a_1\,b} \, =\,
\,1\,+\,g\,G_{x y}^{+}\, v_{+ y}\,
\eeq
and correspondingly
\beq\label{B12}
O^{T}_{x}\,=\,\,1\,+\,g\,v_{+ y}\,G_{y x}^{+}\,,
\eeq
which is redefinition of the operator expansions used in \cite{LipatovEff} in terms of Green's function instead 
integral operators, see Appendex B above.
The Green's function in above equations we understand as Green's function of the $D_{+}$ operator
and express it in the perturbative sense as :
\beq\label{B13}
G_{x y}^{+}\,=\,G_{x y}^{+\,0}\,+\,g\,G_{x z}^{+\,0}\, v_{+ z}\,G_{z y}^{+}\,
\eeq
and
\beq\label{B14}
G_{y x}^{+}\,=\,G_{y x}^{+\,0}\,+\,g\,G_{y z}^{+}\, v_{+ z}\,G_{z x}^{+\,0}\,,
\eeq
with the bare propagators defined as (there is no integration on index $x$ in expressions)
\beq\label{B15}
\D_{+ x}\,\,G_{x y}^{+\,0}\,=\,\delta_{x\,y}\,,\,\,\,G_{y x}^{+\,0}\,\overleftarrow{\D}_{+ x}\,=\,-\delta_{x\,y}\,.
\eeq
The following properties of the operators now can be derived:
\begin{enumerate}
\item
\beqar\label{B161}
\delta\,G^{+}_{x y}& = & g\,G_{x z}^{+\,0}\,\Le \delta v_{+ z} \Ra\,G_{z y}^{+}+
\,G_{x z}^{+\,0}\, v_{+ z}\,\delta G_{z y}^{+}=
g\,G_{x z}^{+\,0}\,\Le \delta v_{+ z} \Ra\,G_{z y}^{+}+
\,G_{x z}^{+\,0}\, v_{+ z}\,\Le \delta G_{z p}^{+} \Ra\,D_{+ p}\,G^{+}_{p y}=\nonumber \\
&=&
g \Le
G_{x z}^{+\,0}\,\Le \delta v_{+ z} \Ra\,G_{z y}^{+}-
G_{x z}^{+\,0}\, v_{+ z}\,G_{z p}^{+}\,\Le \delta D_{+ p}\Ra\,G^{+}_{p y}\Ra=
g \Le
G_{x p}^{+\,0}\,+\,G_{x z}^{+\,0}\, v_{+ z}\,G_{z p}^{+}\Ra\,\delta v_{+ p} \,G^{+}_{p y}=\nonumber \\
&=&\,
\,g\,G^{+}_{x p}\,\delta v_{+ p} \,\,G^{+}_{p y}\,;
\eeqar
\item
\beq\label{B16}
\delta\,O_{x}\, = \,g\,G^{+}_{x y}\,\Le \delta v_{+ y} \Ra\,+g\,\Le \delta G^{+}_{x y}\Ra\,v_{+ y}\,=
\,g\,G^{+}_{x p}\,\delta v_{+ p}\,\Le 1\, +\,g \,G^{+}_{p y}\,v_{+ y}\,\Ra\,=\,
g\,G^{+}_{x p}\,\delta v_{+ p}\,O_{p}\,;
\eeq
\item
\beq\label{B17}
\D_{+ x}\,\delta\,O_{x}\,=\,g\,\Le \D_{+ x}\,G^{+}_{x p} \Ra\,\delta v_{+ p}\,O_{p}\,=\,
g\,\Le 1\,+\,g\,\,v_{+ x}\,G^{+}_{x p}\,\Ra\,\delta v_{+ p}\,O_{p}\,=\,
g\,O^{T}_{x}\,\delta v_{+ x}\,O_{x}\,;
\eeq
\item
\beq\label{B18}
\D_{+ x}\,O_{x}\,=\,g\,\Le \D_{+ x}\,G^{+}_{x y} \Ra\,v_{+ y}\,=\,
g\,v_{+ x}\,\Le 1\,+\,g\,G_{x y}^{+}\,v_{+ y}\,\Ra\,=\,
g\,v_{+ x}\,O_{x}\,;
\eeq
\item
\beq\label{B19}
O_{x}^{T}\,\overleftarrow{\D}_{+ x}\,=\,g\,v_{+ y}\,\Le G^{+}_{y x}\,\overleftarrow{\D}_{+ x} \Ra\,=\,
-\,g\,\Le 1\,+\,v_{+ y}\,\,G^{+}_{y x}\,\Ra\,v_{+ x}\,=\,-g\,O^{T}_{x}\,v_{+ x}\,.
\eeq
\end{enumerate}
We see, that the operator $O$ and $O^{T}$ have the properties of ordered exponents. For example, choosing bare propagators as
\beq\label{B110}
\,G_{x y}^{+\,0}\,=\,\theta(x^{+}\,-\,y^{+})\,\delta^{3}_{x y}\,,\,\,\,
\,G_{y x}^{+\,0}\,=\,\theta(y^{+}\,-\,x^{+})\,\delta^{3}_{x y}\,,\,\,\,
\eeq
we immediately reproduce:
\beq\label{B111}
O_{x}\,=\,P\, e^{g\int_{-\infty}^{x^{+}}\,dx^{'+}\, v_{+}(x^{'+})} \,,\,\,\,
O^{T}_{x}\,=\,P\, e^{g\int_{x^{+}}^{\infty}\,dx^{'+}\, v_{+}(x^{'+})} \,.
\eeq
The form of the bare propagator
$
\,G_{x y}^{+\,0}\,=\,\frac{1}{2}\,\left[\,\theta(x^{+}\,-\,y^{+})\,-\,\theta(y^{+}\,-\,x^{+})\,\right]\,\delta^{3}_{x y}\,
$ 
will lead to the
more complicated representations of $O$ and $O^{T}$ operators, see in \cite{LipatovEff} and \cite{Our4}.
We note also that the Green's function notation $\tilde{G}^{\pm\,0}_{x^{\pm} y^{\pm}}$ in the paper is used for the designation of the only theta function part of the full $G^{\pm 0}_{x^{\pm} y^{\pm}}$ Green's function.

 Now we consider a variation of the action's full current :
\beq\label{B112}
\delta\, tr[v_{+ x}\,O_{x}\,\D_{i}^{2}\,A^{+}]=
\frac{1}{g}\,\delta\, tr[\Le \D_{+ x}\,O_{x} \Ra \D_{i}^{2}\,A^{+}]=\frac{1}{g}\, tr[
\Le\D_{+ x} \delta \,O_{x} \Ra \D_{i}^{2}\,A^{+}] = tr[
O^{T}_{x}\,\delta v_{+ x}\,O_{x}\Le \D_{i}^{2}\,A^{+}\Ra]\,,
\eeq 
which can be rewritten in the familiar form used in the paper:
\beq\label{B1121}
\delta\,\Le v_{+}\,J^{+} \Ra\,=\delta\,tr[\, \Le\, v_{+ x}\,O_{x}\,\D_{i}^{2}\,A^{+}\,\Ra\,]\,=\,-\,
\delta v_{+}^{a}\,tr[\,T_{a}\,O\,T_{b}\,O^{T}\,]\,\Le \D_{i}^{2} A^{+}_{b}\Ra\,.
\eeq
We also note, that with the help of \eq{A11} representation of the $O$ operator
the full action's  current can we written as follows
\beq\label{B113}
 tr[\Le v_{+ x}\,O_{x}\,-\,A_{+}\Ra\,\D_{i}^{2}\,A^{+}\,]\,=\,
tr[\Le v_{+}\,-\,A_{+} + v_{+ x}\,G^{+}_{x y}\,v_{+ y}\,\Ra\,\Le \D_{i}^{2} A^{+}\Ra]\,.
\eeq 

\newpage
\section*{Appendix C: NLO vertex of interactions of reggeized gluons}
\renewcommand{\theequation}{C.\arabic{equation}}
\setcounter{equation}{0}

 The NLO one-loop vertex of reggeized gluons interactions is defined in the formalism as
\beqar\label{C1}
&-&2\,\imath\, K^{a\,b}_{x\,y\,1}\,=\,\Le\,\frac{\delta^{2}\,\ln\Le 1 + G_0\,M\,\Ra}{\delta A_{+\,x}^{a}\,\delta A_{-\,y}^{b}}\,\Ra_{A_{+},\,A_{-},\,v_{f\,\bot}\,=\,0}\,=\,\nonumber \\
&=&\left[ G_0\,\frac{\delta^{2}\,M}{\delta A_{+\,x}^{a} \delta A_{-\,y}^{b}} \Le 1 + G_0\,M\,\Ra^{-1}-
 G_0\,\frac{\delta\,M}{\delta A_{-\,y}^{b}}\Le 1 + G_0\,M\Ra^{-1}
 G_0\,\frac{\delta\,M}{\delta A_{+\,x}^{a}}\Le 1 + G_0\,M\Ra^{-1}
\right]_{A_{+},A_{-},v_{f\,\bot} =\, 0 }\,
\eeqar
where the trace of the expression is assumed. With the help of \eq{Ef5}, see also \cite{Our2}, we have correspondingly:
\beq\label{C2}
-2\,\imath\, K^{a\,b}_{x\,y\,1}\,=\,
\left[ G_0\,\frac{\delta^{2}\,M}{\delta A_{+\,x}^{a} \delta A_{-\,y}^{b}} -
 G_0\,\frac{\delta\,M}{\delta A_{-\,y}^{b}}
 G_0\,\frac{\delta\,M}{\delta A_{+\,x}^{a}}
\right]_{A_{+},\,A_{-},\,v_{f\,\bot} =\, 0 }\,.
\eeq
Taking into account the asymptotically leading contributions of $g^2$ order, that means the $M_{L}$ term presence in the expressions, see \cite{Kovner, Our2}, we obtain:
\beq\label{C3}
-2\,\imath\, K^{a\,b}_{x\,y\,1}\,=\,
\left[ G_0\,\frac{\delta^{2}\,M_{L}}{\delta A_{+\,x}^{a} \delta A_{-\,y}^{b}} -
G_0\,\frac{\delta\,M_{L}}{\delta A_{-\,y}^{b}}G_0\,\frac{\delta\,M_1}{\delta A_{+\,x}^{a}}-
G_0\,\frac{\delta\,M_{1}}{\delta A_{-\,y}^{b}}G_0\,\frac{\delta\,M_L}{\delta A_{+\,x}^{a}}
\right]_{A_{+},\,A_{-},\,v_{f\,\bot} =\, 0 }\,.
\eeq
For the first term we have:
\beq\label{C4}
-2\,\imath\, K^{a\,b}_{x\,y\,1,\,1}\,=\,G_0\,\frac{\delta^{2}\,M_{L}}{\delta A_{+\,x}^{a} \delta A_{-\,y}^{b}}\,=\,
G_{0\, + +}^{t z}\,\frac{g}{N}\,
\frac{\delta\,\Le U_{1}^{c d c}\Ra^{+}_{z t}}{\delta A_{+\,x}^{a}}\,
\frac{\delta\,  \D_{i}^{2} A_{-\,z}^{d}}{\delta A_{-\,y}^{b}}\,
\eeq
where the following identity was used:
\beq\label{C5}
\D_{i}\,\D_{-}\,\rho_{a}^{i}\,=\,-\,\frac{1}{N}\,\D_{i}^{2}\,A_{-}^{a}\,,
\eeq
see \eq{Ef5} and \cite{Our2}.
Using the following expressions
\beq\label{C6}
\frac{\delta\,\Le U_{1}^{c d c}\Ra^{+}_{z t}}{\delta A_{+\,x}^{a}}\,=\,g\,
\Le U_{2}^{c d c a_1}\Ra^{+ +}_{z t w}\,\frac{\delta\, v_{+\,w}^{a_1\,cl}}{\delta\,A_{+\,x}^{a}}\,
\eeq
and
\beq\label{C7}
\frac{\delta\, v_{+\,w}^{a_1\,cl}}{\delta\,A_{+\,x}^{a}}\,=\,\delta^{a\,a_1}\,\Le \delta^{2}_{x_{\bot}\,w_{\bot}} \delta_{x^{+}\,w^{+}}\Ra\,
\eeq
to requested accuracy, we obtain for \eq{C4}:
\beq\label{C8}
-2\,\imath\, K^{a\,b}_{x\,y\,1,\,1}\,=\,\frac{g^2}{N}\,G_{0\, + +}^{t z}\,\Le U_{2}^{c b c a_1}\Ra^{+ +}_{z t w}\,
\Le \delta^{a\,a_1}\,\delta^{2}_{x_{\bot}\,w_{\bot}} \delta_{x^{+}\,w^{+}}\,  \Ra\,
\Le \delta^{2}_{y_{\bot}\,z_{\bot}} \delta_{y^{-}\,z^{-}} \D_{i\,z}^{2}\,\Ra\,,
\eeq
where the NNLO term of the Lipatov's current series expansion reads as
\beqar\label{C9}
\Le\Le U_{2}^{c bc a}\Ra^{+ +}_{z t w}\,\Ra_{A_{+},\,A_{-} = \, 0}\,& = &\,\frac{1}{2}\,N^{2}\,\delta^{a b}\,\left[\,\Le\,
G_{z w}^{+\,0}\,G_{w t}^{+\,0}\,+\,G_{t w}^{+\,0}\,G_{w z}^{+\,0}\,\Ra\,+\,\right.\nonumber \\
&+&\,\left.
\,2\,\Le
G_{z t}^{+\,0}\,G_{t w}^{+\,0}\,+\,G_{z t}^{+\,0}\,G_{w z}^{+\,0}\,+\,G_{z w}^{+\,0}\,G_{t z}^{+\,0}\,+\,G_{t z}^{+\,0}\,G_{w t}^{+\,0}\,
\Ra\,\right]\,.
\eeqar
Therefore, writing explicitly all integrations in the expression, we obtain:
\beqar\label{C10}
&\,& -\,2\,\imath\, K^{a\,b}_{x\,y\,1,\,1}\,  = 
 \frac{1}{2}\, g^{2}\,N\,\delta^{a\,b}\,\int d^{4} z\,d^{4} t\, d^{4} w\,\Le\,\Le \D_{i\,z}^{2}\,G_{0\,+ +}^{t z}\Ra\, 
\Le \delta^{2}_{x_{\bot}\,w_{\bot}} \delta_{x^{+}\,w^{+}}\Ra\,\Le \delta^{2}_{y_{\bot}\,z_{\bot}} \delta_{y^{-}\,z^{-}}\Ra\,\right.\cdot\nonumber \\
&\cdot& \left. \left[\Le
G_{z w}^{+\,0}\,G_{w t}^{+\,0}+G_{t w}^{+\,0}\,G_{w z}^{+\,0}\Ra+
2\Le
G_{z t}^{+\,0}\,G_{t w}^{+\,0}+G_{z t}^{+\,0}\,G_{w z}^{+\,0}+G_{z w}^{+\,0}\,G_{t z}^{+\,0}+G_{t z}^{+\,0}\,G_{w t}^{+\,0}
\Ra\right] \Ra.
\eeqar

 Formally, there are three additional terms are present in \eq{C2}. The first one
\beq\label{C11}
-2\,\imath\, K^{a\,b}_{x\,y\,1,\,2}\,=\,-\,G_{0\, + +}\,\frac{\delta\,M_{L}}{\delta A_{-\,y}^{b}}\,\,G_{0\, + i}\,
\frac{\delta\,M_{1\,i -}}{\delta A_{+\,x}^{a}}\,,
\eeq
the second one
\beq\label{C12}
-2\,\imath\, K^{a\,b}_{x\,y\,1,\,3}\,=\,-\,G_{0\, i +}\,\frac{\delta\,M_{L}}{\delta A_{-\,y}^{b}}\,\,G_{0\, + +}\,
\frac{\delta\,M_{1\,- i}}{\delta A_{+\,x}^{a}}\,
\eeq
and the third one
\beq\label{C13}
-2\,\imath\, K^{a\,b}_{x\,y\,1,\,4}\,=\,-\,G_{0\, i +}\,\frac{\delta\,M_{L}}{\delta A_{-\,y}^{b}}\,G_{0\, + j}\,
\frac{\delta\,M_{1\,j i}}{\delta A_{+\,x}^{a}}\,.
\eeq
Nevertheless, only the third one contributes to the kernel in the limit of zero Reggeon fields, we have there: 
\beq\label{C14}
\frac{\delta\,M_{L}^{c d}}{\delta A_{-\,y}^{b}}\,=\,\frac{g}{N}\,
\Le U_{1}^{c b d}\Ra^{+}_{t w}\,\Le \delta^{2}_{y_{\bot}\,t_{\bot}} \delta_{y^{-}\,t^{-}} \D_{i\,t}^{2}\,\Ra\,
\eeq
where
\beq\label{C15}
\Le\Le U_{1}^{c b d}\Ra^{+}_{t w}\Ra_{A_{+},\,A_{-},\,v_{f\,\bot} =\, 0 }\,=\,\frac{1}{2}\,N\,f_{c d b}\,\Le G^{+\,0}_{t w}\,-\,G^{+\,0}_{w t} \Ra\,.
\eeq
Also we have:
\beq\label{C16}
\frac{\delta\,M_{j i}^{d c}}{\delta A_{+\,x}^{a}}\,=\,2\,g\,f_{d a c}\,\delta_{j\,i}\,\delta^{2}_{z_{\bot}\,x_{\bot}}\,\delta_{z^{+}\,x^{+}}\,\D_{-\,z}\,.
\eeq
The final expresion for this terms reads, therefore, as:
\beq\label{C17}
-2\imath\, K^{a\,b}_{x\,y\,1,\,4}=-\,g^2\,N\,\delta^{a b}\,\int\,d^4t\, d^4w\, d^4 z \,
\Le G^{+\,0}_{t w}-G^{+\,0}_{w t} \Ra\,\delta^{2}_{z_{\bot}\,x_{\bot}}\, \delta_{z^{+}\,x^{+}}\,\delta^{2}_{y_{\bot}\,t_{\bot}}\, \delta_{y^{-}\,t^{-}} \,
\Le\,G_{0\, + i}^{w z}\,\D_{-\,z}\,\D_{i\,t}^{2}\,G_{0\, i +}^{z t}\Ra\,.
\eeq
We notice that both \eq{C17} and \eq{C10} contributions are precisely the same as obtained in \cite{Our2} paper. Therefore, we immediately write the full contribution
from \cite{Our2} which is
\beq\label{C18}
K^{a\,b}_{x\,y\,1}\,= \,-\,\frac{\,g^{2}\,N}{8\,\pi}\,\D_{i\,x}^{2}\,\Le\,\int\,\frac{d p_{-}}{p_{-}}\,\int\,\frac{d^{2} p_{\bot}}{(2 \pi)^{2}}\,\int\,\frac{d^{2} k_{\bot}}{(2 \pi)^{2}}\,
\frac{ \,k_{\bot}^{2}}{p_{\bot}^{2}\,\Le\,p_{\bot}\,-\,k_{\bot}\,\Ra^{2}}\,e^{-\imath\,\,k_{i} \,\Le x^{i}\,-\,y^{i}\Ra}\,\Ra\,.
\eeq
We can rewrite this expression redefining the vertex in \eq{Pro6}  as
\beq\label{C19}
K^{a\,b}_{x\,y\,1}\,\rightarrow\,K^{a\,b}_{x\,y\,1}\,\D_{i\,x}^{2}\,= \,\int\,\frac{d^{2} p}{(2 \pi)^2}\,\tilde{K}(p)\,e^{-\imath\,p_i\,\Le x^{i}\, - \,y^{i} \Ra}\,\D_{i\,x}^{2}\,
\eeq
with
\beq\label{C20}
\tilde{K}(p \,,\,\eta)\,=\,-\,\frac{\,N\,\pi\,g^{2}}{2}\,\delta(p_{\,+})\,\delta(p_{\,-})\,
\int_{0}^{\eta} d \eta^{'}\,\int \frac{d^2 k_{\bot}}{(2 \pi)^2}\,\frac{p_{\bot}^2}{k_{\bot}^{2}\,\Le p_{\bot} - k_{\bot} \Ra^2}\,,
\eeq
where the physical cut-off $\eta$ in rapidity space $y\,=\,\frac{1}{2} \ln(\Lambda\,k_-)$ is introduced.

\newpage
\section*{Appendix D: calculation of \eq{Corr7} integral}
\renewcommand{\theequation}{D.\arabic{equation}}
\setcounter{equation}{0}

 The \eq{Corr7} answer can be obtained immediately  if we will note that \eq{Corr21} expression is not symmetrized in respect with $a_4$ and $a_1$ color indexes. Indeed, we can use the same expression
with the indexes permutated (we also can use fully symmetrized expression) obtaining immediately \eq{Corr10} answer. Nevertheless, it is instructive to calculate the first r.h.s. term of \eq{Corr7} directly, we have for the next to leading order terms in the integral:
\beq\label{D1}
I\,=\,\frac{4^{1+\varepsilon}}{\pi^{1-\varepsilon}}\,\sum_{n=2}^{\infty}\,\frac{(-1)^{n-1}\,Y^n}{n!}\,\Le \frac{\Le N\,\alpha_{s}\Ra^{n}}{(4\pi)^{n(1+\varepsilon)}}\Ra\,\Gamma^{\,n-1}(1-\varepsilon)\,\Le \frac{2}{\varepsilon}\Ra^{n-1}\,
\int \frac{d^{2} k_{\bot}}{\Le k_{\bot}^{2}\Ra^{1-\lambda}\,(p_{\bot}\,-\,k_{\bot})^{2}}\,,
\eeq
with $\lambda\,=(n-1)\varepsilon$. Now we use standard formulas:
\beq\label{D2}
 \int d^{D} k\,\frac{1}{k_{\bot}^{2 a}(p_{\bot}\,-\,k_{\bot})^{2 b}}\,=\,\frac{(2 \pi)^{D}}{(4 \pi)^{D/2}}\,
\frac{\Gamma(D/2 -a)\,\Gamma(D/2 - b)\,\Gamma(a+b-D/2)}{\Gamma(a)\,\Gamma(b)\,\Gamma(D- a -b)}\,\frac{1}{(p_{\bot}^{2})^{a+b-D/2}},
\eeq
Taking $D\,=\,2\,+\,2\,\varepsilon_{1}$ and taking at the end $\varepsilon_1 \rightarrow \varepsilon $ we see that the obtained answer is proportional to
\beq\label{D3}
\frac{\Gamma(\varepsilon_{1}+\lambda)\,\Gamma(\varepsilon_1)\,\Gamma(1-\varepsilon_1-\lambda)}{\Gamma(1-\lambda)\Gamma(2\varepsilon_{1}+ \lambda)}\,\propto\,
\frac{1}{\varepsilon}\,\Le 1+\frac{1}{n}\Ra\,+\,\gamma_{E}\,\Le 1+\frac{1}{n}\Ra\,.
\eeq
Therefore we obtain for the \eq{D1} sum:
\beq\label{D4}
I=-\frac{4^{1+\varepsilon}}{p_{\bot}^{2}\pi^{-2\varepsilon}}\frac{\varepsilon}{2\,\Gamma(1-\varepsilon)}
\sum_{n=2}^{\infty}\,\frac{(-1)^{n} Y^n}{n!}\Le \frac{\Le N\,\alpha_{s}\Ra^{n}}{(4\pi)^{n(1+\varepsilon)}}\Ra\Gamma^{\,n}(1-\varepsilon)\Le \frac{2}{\varepsilon}\Ra^{n}
\Le \frac{1}{\varepsilon}\,\Le 1+\frac{1}{n}\Ra+\gamma_{E}\Le 1+\frac{1}{n}\Ra \Ra\Le p_{\bot}^{2}\Ra^{n\varepsilon}\,.
\eeq
We note, that the answer for the \eq{Corr7} propagator reproduces \eq{Corr9} leading order expression if we will take  $n=1$ in the sum, therefore the full answer can be obtained 
by expanding the summation in the expression to $n=1$:
\beqar\label{D44}
\tilde{{\cal G}}^{a b}_{+}&\propto&\, -\delta^{a b}\,\frac{4^{1+\varepsilon}}{p_{\bot}^{2}\pi^{-2\varepsilon}}\frac{\varepsilon}{2\,\Gamma(1-\varepsilon)} \, \\
&\,&
\sum_{n=1}^{\infty}\,\frac{(-1)^{n} Y^n}{n!}\Le \frac{\Le N\,\alpha_{s}\Ra^{n}}{(4\pi)^{n(1+\varepsilon)}}\Ra\Gamma^{\,n}(1-\varepsilon)\Le \frac{2}{\varepsilon}\Ra^{n}
\Le \frac{1}{\varepsilon}\,\Le 1+\frac{1}{n}\Ra+\gamma_{E}\Le 1+\frac{1}{n}\Ra \Ra\Le p_{\bot}^{2}\Ra^{n\varepsilon}= \nonumber \\
&=&\,-\,\frac{2\,\delta^{a b}}{p_{\bot}^{2}}\,\Le \sum_{n=1}^{\infty}\,\frac{\Le\epsilon(p_{\bot}^{2})\,Y\Ra^{n}}{n!}\,+\,\sum_{n=1}^{\infty}\,\frac{\Le\epsilon(p_{\bot}^{2})\,Y\Ra^{n}}{n\,n!}\,\Ra\,=\,
\frac{2\,\delta^{a b}}{p_{\bot}^{2}}\,\Le 1\,-\,e^{\epsilon(p_{\bot}^{2})\,Y}\,-\,\int_{0}^{-\,\epsilon(p_{\bot}^{2})\,Y}\,\frac{d y}{y}\Le e^{-y}\,-\,1\,\Ra \Ra\,. \nonumber
\eeqar

\newpage
\section*{Appendix E: LO target-projectile symmetry and light-cone gauge}
\renewcommand{\theequation}{E.\arabic{equation}}
\setcounter{equation}{0}

\begin{figure}[!h]
\centering
\includegraphics[scale=.9]{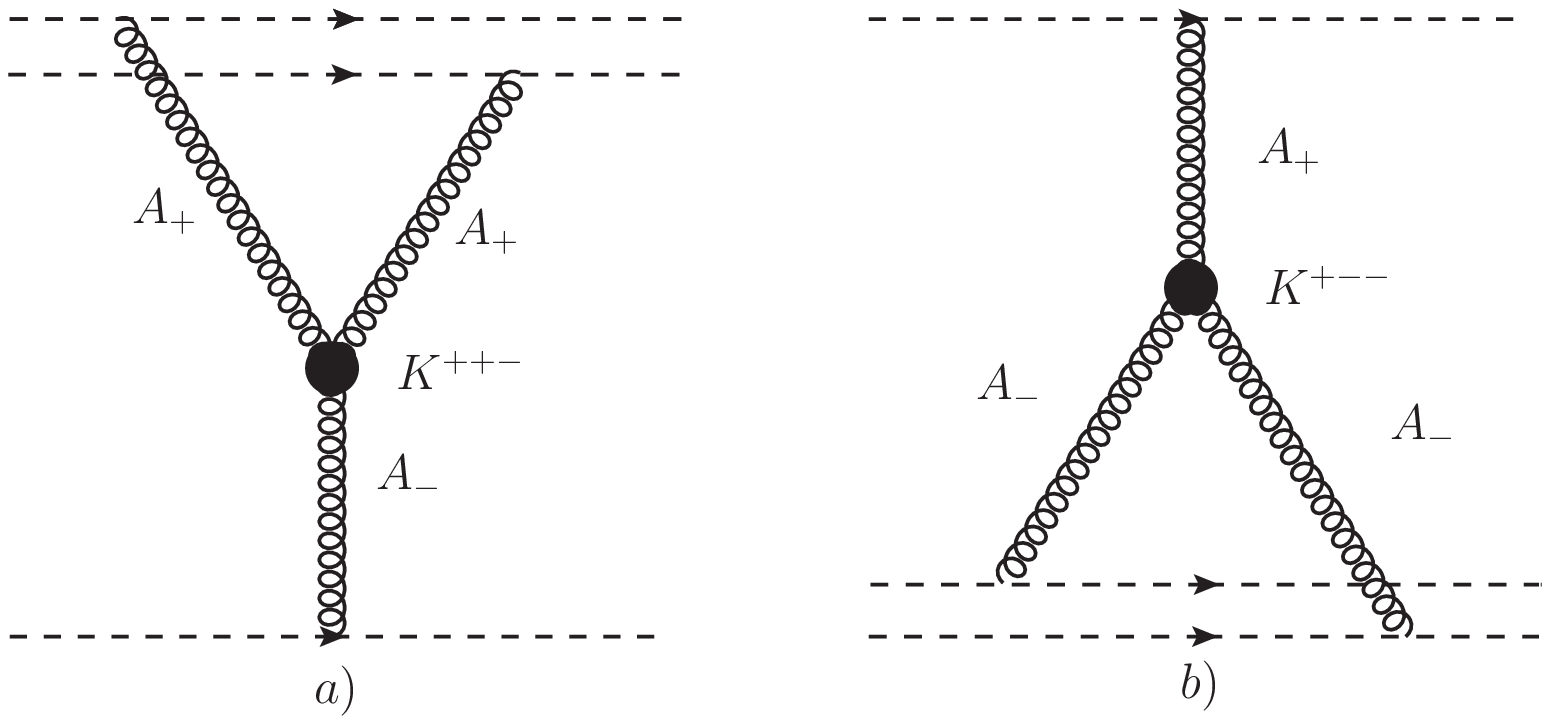}
\caption{The diagrams represent simplest scattering processes with triple Reggeon vertices involved. }
\label{d}
\end{figure}

 Let's consider the simplest scattering process of Reggeons with triple vertices included, here are two close in rapidity space particles  are scattering off the third one in the process
described by the multi-Regge kinematics regime. In this case there are two diagrams \fig{d}-a and \fig{d}-b which must be equal if we request the target-projectile symmetry (change of $+$ to $-$) and gauge independence of the amplitudes\footnote{In the covariant gauges the symmetry is preserved from the beginning, therefore demonstrating the target-projectile symmetry 
for the diagram we demonstrate the gauge independence of the calculation results, the case when the gauge choice changes the overall diagram's coefficient is not assumed.}. Further we assume, that the framework correctly reproduces the mathematical structure of the 
amplitudes, i.e. we have to check only the equivalence of the numerical coefficients of the diagrams where the correlators are convoluted with the correspondence impact factors. 
The final answer for the presented diagrams, therefore, depends on the coefficients in the front of the known vertices \eq{BV19}-\eq{BV20} and coefficients of the unknown impact factors, which 
structure in this case is simple of course. Namely, we consider the scattering of the gluons and the vertices are simply proportional to $f^{a b c}$ structure constant with unknown yet numerical coefficients. 
So, first of all, we write to which coefficients the diagrams are proportional, the \fig{d}-a diagram is proportional to
\beq\label{E1}
I_{a)}\,\propto\,\frac{1}{2}\,V_{+}^2\,V_{-}
\eeq
expression, whereas  \fig{d}-b diagram has the following numerical coefficient:
\beq\label{E2}
I_{b)}\,\propto\,-\,V_{+}\,V_{-}^2\,,
\eeq
where the $V_{\pm}$ denote the impact factors of interaction of the corresponding Reggeon fields with external gluons.
The comparison of the diagrams provides in turn the following condition for the impact factors which must be satisfied:
\beq\label{E3}
\,V_{+}\,=\,-\,2\,V_{-}\,.
\eeq
Basing on the \cite{Our3} results we know, that the impact factors  can be defined similarly to the definition of the kernels, see \eq{BV1}, with only change of the 
variation with respect to Reggeon field to the variation with respect to the corresponding on-shell  free field, in our case this is $v_{f\bot}$ gluon fields.
Namely we have
\beq\label{E4}
V_{+}\,\propto\,K^{f\,f\,+}\,,\,\,V_{-}\,\propto\,K^{f\,f\,-}\,.
\eeq
Now we have to examine how the $v_{f\bot}$ fields are arising in the RFT Lagrangian. For the case of $V_{+}$ vertex, these fields arise in the impact factor
only through $v_{\bot}$ fields. Taking into account that to the first leading order the $A_{-}$ Reggeon field appears in the expressions only 
through transverse field 
\beq\label{E5}
v_{\bot}\propto\,v_{f\bot}\,+\,\hat{C}_{\bot}^{-}\,A_{-}\,,
\eeq
here $\hat{C}_{\bot}^{-}$ is some operator, see \cite{Our1}-\cite{Our2}, we conclude that
\beq\label{E6}
V_{+}\,\propto\,K^{f\,f\,+}\,\propto\,K^{--+}\,\propto\,-f^{a b c}\,.
\eeq
For the second impact factor there are two possibilities to appear. The first one is the following one:
\beq\label{E7}
V_{-}\,\propto\,K^{f\,f\,-}\,\propto\,K^{\bot \bot -}\,\propto\,K^{\bot \bot \bot}\,=\,0\,,
\eeq
there are no such vertices in the Lagrangian to LO precision. The second form of the vertex is determined by the
usual dependence of the free longitudinal field on the transverse one in the light cone gauge:
\beq\label{E8}
v_{+}\,\propto\,v_{f +}\,+\,A_{+}\,=\,\hat{C}_{+\bot}\,v_{f\bot}\,+\,A_{+}
\eeq
that provides:
\beq\label{E9}
V_{-}\,\propto\,K^{f\,f\,-}\,\propto\,K^{+ + -}\,\propto\,\frac{1}{2}\,f^{a b c}\,.
\eeq
Finally we see that indeed
\beq
V_{+}\,=\,-2\,V_{-}\,
\eeq
as expected.

\end{document}